\documentclass{aa}
\usepackage[varg]{txfonts}
\usepackage{graphicx}
\usepackage{multirow}
\usepackage{caption}
\usepackage{subcaption}

\usepackage[symbol]{footmisc}

\begin{document}

\title{{\it Gaia} Data Release 3. Astrometric binary star processing}

\author{Jean-Louis Halbwachs\inst{1}
  \and Dimitri Pourbaix\thanks{deceased on 14 November 2021}\inst{2}
  \and Fr\'ed\'eric Arenou\inst{3}
  \and Laurent Galluccio\inst{4}
  \and Patrick Guillout\inst{1}
  \and Nathalie Bauchet\inst{5}
  \and Olivier Marchal\inst{1}
  \and Gilles Sadowski\inst{2}
  \and David Teyssier\inst{6}
  }

\offprints{J.-L. Halbwachs, \email{jean-Louis.halbwachs@astro.unistra.fr}}

\institute{Universit\'e de Strasbourg, CNRS, Observatoire astronomique de Strasbourg, UMR 7550, 
11 rue de l'Universit\'{e}, 67000 Strasbourg, France
\and FNRS, Institut d'Astronomie et d'Astrophysique, Universit\'{e} Libre de Bruxelles, boulevard du Triomphe, 1050 Bruxelles, Belgium
\and GEPI, Observatoire de Paris, Universit\'{e} PSL, CNRS, 5 Place Jules Janssen, 92190 Meudon, France
\and Universit\'e C\^ote d’Azur, Observatoire de la C\^ote d’Azur, CNRS, Laboratoire Lagrange, Bd de l’Observatoire, CS 34229, 06304 Nice Cedex 4, France 
\and IMCCE, Observatoire de Paris, Université PSL, CNRS, Sorbonne Université, Univ. Lille, 77 av. Denfert-Rochereau, 75014, Paris, France
\and Telespazio UK S.L. for ESA/ESAC, Camino bajo del Castillo, s/n, Urbanizacion Villafranca del Castillo, Villanueva de la Cañada, 28692, Madrid, Spain
}

\date{Received/Accepted}

\abstract {The {\it Gaia} Early Data Release 3 contained the positions, parallaxes and proper motions of 1.5 billion sources, among which some did not fit well the "single star" model. Binarity is one of the causes of this.}
{Four million of these stars were selected and various models were tested to detect binary stars and to derive their parameters.}{A preliminary treatment was used to discard the partially resolved double stars and to correct the transits for perspective acceleration. It was then investigated whether the measurements fit well with an acceleration model with or without jerk. The orbital model was tried when the fit of any acceleration model was beyond our acceptance criteria. A Variability-Induced Mover (VIM) model was also tried when the star was photometrically variable. A final selection has been made in order to keep only solutions that probably correspond to the real nature of the stars.}
{At the end, $338\,215$ acceleration solutions, about $165\,500$ orbital solutions and 869 VIM solutions were retained. In addition, formulae for calculating the uncertainties of the Campbell orbital elements from orbital solutions expressed in Thiele-Innes elements are given in an appendix.}{}


\maketitle 

\titlerunning{{\it Gaia} DR3: Binary stars}
\authorrunning{Halbwachs et al.}
\renewcommand*{\thefootnote}{\arabic{footnote}}


\section{Introduction}

\subsection{Context}

Since the start of the {\it Gaia} satellite observations in summer 2014, several data releases have been made available to the scientific community.
The astrometric parameters of more than one billion stars have been determined three times with ever increasing precision, but these data still 
described the stars as single. The third Data Release (DR3) and the Early third Data Release (EDR3) are the first to have a significant number of observations covering a sufficiently 
long time span (about 33 months) to make it worthwhile to use {\it Gaia} observations to search for binary stars. 

The extension of the exploitation of {\it Gaia} observations to binary stars reveals two more difficulties than the application of the single star model. 
As with its predecessor Hipparcos, the double star candidates are covered by different numerical models, without being certain that any of them really 
corresponds to a given star. Therefore, we have to solve the problem of choosing the model that best applies to a given star, as well as the 
final acceptance of the solution. Before that, we have to filter the input data to retain only those stars that may be able to get a suitable solution.


\subsection{Selection of stars to be processed}

The astrometric processing of the {\it Gaia} Early Data Release 3 \citep[EDR3,][]{EDR3} left some 36.5 million stars that were considered to be possibly double.
These stars were all brighter than G=19, had a renormalized unit weight error \citep[$RUWE$; see][]{Lindegren21} greater than 1.4, and were observed over
at least 12 visibility periods. These objects included many pairs of stars that were resolved whenever the orientation of the scan direction allowed, and seen as a single photocentre otherwise. These so-called ``partially resolved double stars'' are likely to appear as unresolved binary stars of extremely rapid orbital motion, and should be discarded as far as possible. It is planned to apply a specific treatment to them in the next DRs, but for the moment we can only eliminate them when possible. For this purpose, two quantities published in {\it Gaia} EDR3 and coming from the astrometric binary detection processing \citep{Lindegren21} were used to establish two additional criteria: 
The percentage of CCD observations where image parameter determination (IPD) detected more than one peak, called
$ipd\_frac\_multi\_peak$, must be less than or equal to 2, and $ipd\_gof\_harmonic\_amplitude$, which is the amplitude of the natural logarithm of the goodness-of-fit obtained in the IPD versus position angle of scan, must be less than 0.1.
As a result, the number of targets to be treated was reduced to 10.9 million. 
In order to further reduce the number of partially resolved double stars, a final selection was based on photometric properties. 
As the pixels of the astrometric field are much smaller than those of the BP and RP windows, a component of a close binary
at the resolution limit appears less bright in $G$ magnitude (from the astrometric field) than one would expect from its magnitude in the BP and RP fields.
The corrected BP and RP flux excess factor $C^*$ and its uncertainty $\sigma_{C^*}$ as defined in \citet[Eq.~6 and 18]{Riello21} were taken into account. The drastic condition
$|C^*| < 1.645 \; \sigma_{C^*}$,
corresponding to a 90\% confidence interval, was applied and a selection of $4\,115\,743$ stars was finally obtained. 


\section{Treatment overview}

\begin{figure}
\centering
\includegraphics[width=9cm]{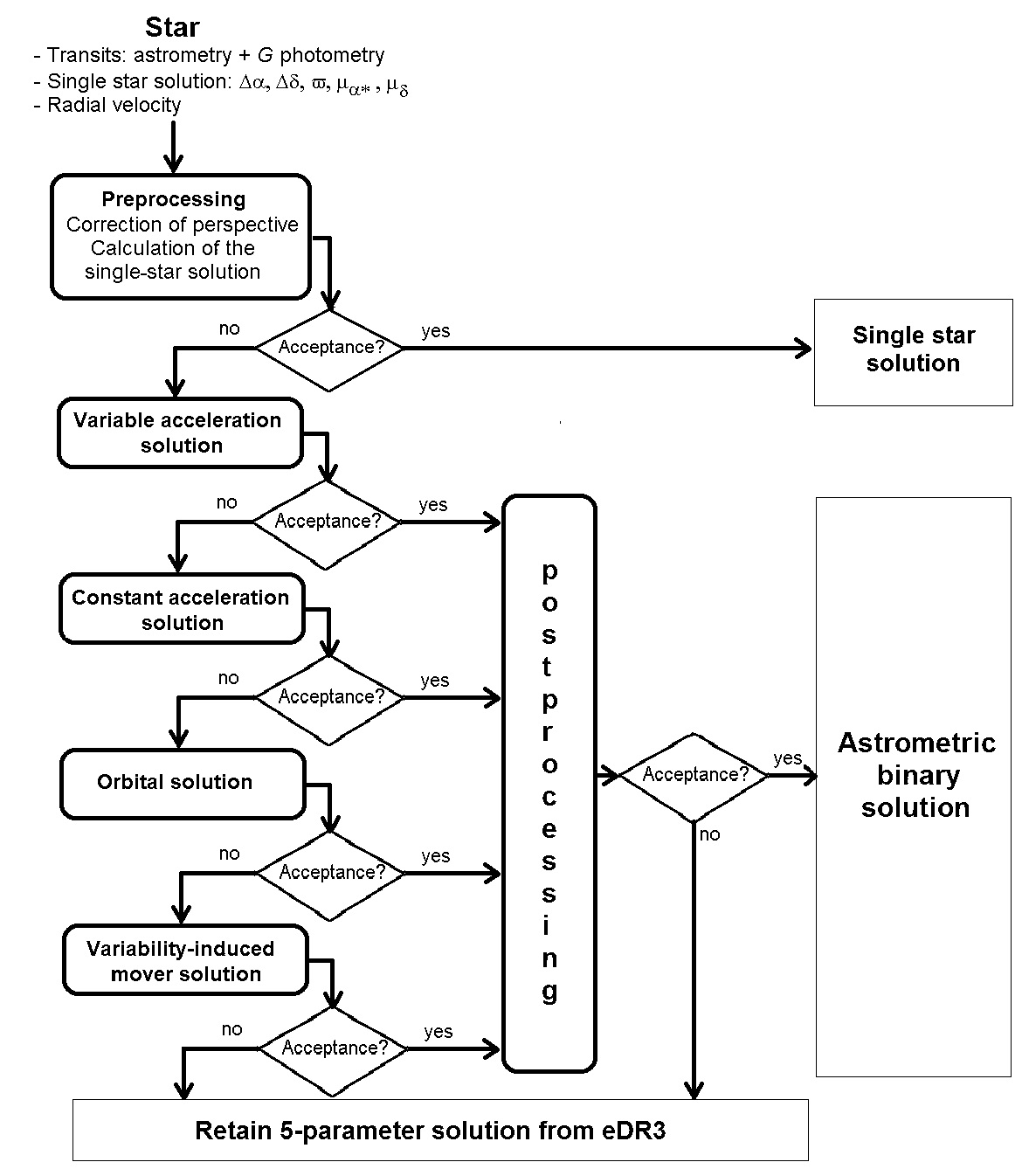}
\caption{Overall organisation of the astrometric treatment of binary stars, as it was eventually applied to the DR3. 
The cascade on the left side of the figure is the so-called ``main processing'' hereafter.}
\label{fig:over}
\end{figure}

\subsection{Overall organisation}
\label{sect:organisation}

The organisation of the astrometric binary star processing, as it was finally realised, is summarised in Fig.~\ref{fig:over}. 
The input data consist of the $4\,115\,743$ selected stars, for which the following information is used:
the along-scan (AL) coordinates of each transit of the astrometric field of view, the partial derivatives of the single star astrometric model, the
$G$ magnitudes of the transits, and the parameters of a preliminary single star solution, i.e. the coordinate offsets, the parallax and the
proper motion. The production of the astrometric data is described in \citet{Lindegren21}, while the magnitude reduction is explained in
\citet{Riello21}. The data are complemented by radial velocities, when available \citep{Seabroke21}.

The calculation of the astrometric solutions is done in the so-called main processing, as explained hereafter.
The data discussed above are ingested by the preprocessing, which prepares them for the calculation of different types of solutions.
The solutions are calculated one after the other, starting with a new calculation of the single star solution, then the acceleration solutions,
the orbital solution, and, finally, the variability-induced mover (VIM) solutions if the star is photometrically variable. 
As soon as a solution is considered acceptable, the processing cascade is interrupted and the star goes to post-processing.
The post-processing essentially consists of re-filtering the solutions, applying more stringent criteria than those used after the solutions were
calculated. When a solution is discarded by post-processing, it is replaced by the EDR3 single star solution.

In practice, the software for managing the cascade of calculation and acceptance of solutions is much more sophisticated than the scheme shown in
Fig.~\ref{fig:over}. In fact, the solution that entered the post-processing was sometimes a so-called ``alternative'' solution \citep[not to be confused with the "OrbitalAlternative" solutions in][]{Holl}: 
a solution that met poor acceptance criteria, and which was retained because it was the solution that best matched the observations at the end of the cascade. More information on this aspect of the process is given at the end of Section~\ref{subsubsec:signif}, but we can already highlight two things: First, that we did not systematically try all the models to retain the one that was the most suitable; this was motivated mainly by the concern to save computational time, but also by the fear of accepting complex solutions for objects that could fall under a simpler model (for example, giving an acceleration solution for an orbital binary is a less serious error than giving a false orbital solution for a binary that should only have an acceleration solution). 
Despite its apparent mathematical simplicity, the VIMF model has been placed at the end because it combines data of two kinds (astrometric and photometric), which undermines the reliability of the solutions.
Secondly, post-processing consists only of rejecting solutions, without trying to replace them with more complex or alternative solutions. Ideally, the filters applied in post-processing should be included in the main processing, but this was not possible in the time available.


\subsection{Criteria for the acceptance of solutions}

We define hereafter two quality criteria, the goodness of fit and the significance, that can distinguish 
acceptable solutions from those that are irrelevant or at least potentially wrong. In order to decide whether a 
solution is really plausible, we will also consider astrophysical criteria and other constraints, which will 
allow us to identify some solutions as not very plausible. These bad solutions will allow us to delimit a quality domain where they are absent 
or rare, and whose solutions will be kept.


\subsubsection{The goodness of fit}
\label{susubsec:F2}

The goodness of fit indicates if the model provides predictions which are compatible with the actual measurements.
As in the preparation of the Hipparcos catalogue \citep[vol. 1, Sect. 2.1]{hip}, we use the $F_2$ 
estimator, which is given by the formula:

\begin{equation}
F_2 = \sqrt{\frac{9 \nu}{2}}\left[ \left( \frac{\chi^2}{\nu}\right)^{1/3} + \frac{2}{9 \nu} - 1 \right]
\label{eq:F2}
\end{equation}

where $\nu$ is the number of degrees of freedom, and $\chi^2$ is the sum of the squared normalised residuals. 
For a solution derived through an adequate model, $F_2$ is expected to obey the normal distribution
${\cal N}(0,1)$. This property is attributed to linear models \citep{Kendall}, but we verified by simulations
that it also applies to the orbital model presented hereafter, when the semi-major axis of the orbit is significantly
larger than its uncertainty.
A large $F_2$ means that the model is inadequate, or that the uncertainties used to derive the $\chi^2$ are
underestimated. In practice, we know that this last possibility widely applies to the astrometric transits used in DR3.
Therefore, only very large values of $F_2$ indicate that the model cannot be accepted.
Nevertheless, assuming that the model is adequate, the uncertainties may be revised in order to obtain a zero
$F_2$. The simplest correction method, since it keeps the relative weights of the transits and therefore the solution
of the model, consists in multiplying the uncertainties of the observations by the coefficient:

\begin{equation}
c = \sqrt{\frac{\chi^2}{\nu \left( 1-\frac{2}{9\nu} \right)^3}}
\label{eq:corsig}
\end{equation}

It is worth noticing that the correction by the coefficient $c$ also applies to the uncertainties of the model
parameters. In addition to the correction of uncertainties, $F_2$ was also taken into account for the 
selection of solutions. In the main processing, solutions of $F_2 >1000$, if any, were rejected, 
but a stricter condition was applied in the post-processing:
whatever the model, purely astrometric solutions with uncorrected $F_2>25$ were considered as questionable, and were finally not retained.

Although similar in principle, this approach is slightly different from that followed in the treatment of single star solutions for EDR3
\citep{Lindegren21}: Eq.~\ref{eq:corsig} shows that the coefficient $c$ is equal to $\rm{UWE} \times [1-(2/9\nu)]^{-3/2}$, which is 
approximately equal to $\rm{UWE}$. On the other hand, we do not apply a renormalisation, but this does not matter since we only consider 
stars brighter than $G=19$.


\subsubsection{The significance}
\label{subsubsec:signif}

The significance is a dimensionless quantity that was already introduced in the Hipparcos catalogue 
\citep[vol. 1, Sect. 2.3.3]{hip} in order to decide whether the use of a model including additional parameters
is really justified. When the additional parameters are the coordinates of a two-dimensional vector, as is the case for the acceleration models in Sect.~\ref{sec:acc}, the significance, $s$, is defined as the module of this vector
divided by its uncertainty. Therefore, $s$ is calculated with the equation:

\begin{equation}
s = \frac{1}{\sigma_1 \sigma_2} \sqrt{ \frac{p_1^2 \sigma_2^2 + p_2^2 \sigma_1^2 - 2 p_1 p_2 \rho_{12} \sigma_1 \sigma_2 }
{1 - \rho^2_{12}} }
\label{eq:signif}
\end{equation}

\noindent
where $p_1$ and $p_2$ are the coordinates of the additional vector characterising the model (for instance, the acceleration),
$\sigma_1$ and $\sigma_2$ are the uncertainties of $p_1$ and $p_2$, respectively, corrected by the coefficient $c$ derived
with Eq.~\ref{eq:corsig}, and $\rho_{12}$ the correlation coefficient between $p_1$ and $p_2$. The values of
$\sigma_1$, $\sigma_2$ and $\rho_{12}$ are all taken from the variance-covariance matrix.

When the solutions were calculated, a preliminary selection was made through basic filtering: the solutions of $s>12$ and 
$F_2<25$ were considered good enough to be retained, without
trying other models. Otherwise, the solutions with $s>5$ and $F_2 < 1000$ were considered as ``alternative'' solutions (see Sect.~\ref{sect:organisation}), 
and the smaller $F_2$
alternative was provisionally accepted at the end of the cascade. The limit of 5 was adopted empirically on the basis of initial tests of the data. 
Thresholds of 5 and 12, which amount to as many $\sigma$s, seem exceptionally high and are much higher than the thresholds of 3 and 4 that 
were adopted on the basis of simulations for alternative solutions and for direct acceptance respectively. This is because, in reality the uncertainties of 
the transits of a star are certainly not all overestimated by the same coefficient; correcting the uncertainties by a single coefficient is only a stopgap measure,
as it is not possible to separate transits with a correct original uncertainty from those with a very underestimated uncertainty.

The set of solutions thus selected was then subjected to a more severe filtering in order to keep only the most credible solutions, as explained below for each model. The final selection criteria are presented in Table~\ref{tab:threshold}.
For the acceleration models and for the VIMF model, they are more severe than the direct acceptance criteria, and they lead to the rejection of alternative solutions. However, all alternative solutions contribute to the statistical properties of the main processing solutions, which are presented hereafter, and which are the basis for the choice of the final criteria.

\begin{table*}
\caption{Properties of the different models and final conditions for the selection of solutions.}
\begin{tabular}{lrccccc}
\hline
\multirow{2}{*}{Model} & \multirow{2}{*}{Dimension} & \multirow{2}{*}{Significance} & \multicolumn{4}{c}{Selection conditions}      \\
                       &                           &                               & significance & $F_2$  & $\varpi/\sigma_\varpi$ & Other \\
\hline    
&&&&&& \\
Acceleration : &&&&&& \\
&&&&&& \\
Constant  &  7 & $s_7=\frac{g}{\sigma_g}$             & $>20$ & $<22$ & $>1.2 s_7^{1.05}$ & -    \\
Variable  &  9 & $s_9=\frac{\dot{g}}{\sigma_{\dot{g}}}$ & $>20$ & $<25$ & $>2.1 s_9^{1.05}$ & -  \\
&&&&&& \\
\hline    
&&&&&& \\
Orbital$^*$ : &&&&&& \\
eccentric & 12 & $\frac{a_0}{\sigma_{a_0}}$          & $> \mathrm{max} \left( 5,\frac{158}{\sqrt{P_{\mathrm{days}}}} \right)$ & $<25$ & $>\frac{20\;000}{P_{\mathrm{days}}}$  &  $\sigma_e < 0.079 \ln P_{\mathrm{days}} - 0.244$ \\
circular or &  \multirow{2}{*}{10} & \multirow{2}{*}{$\frac{a_0}{\sigma_{a_0}}$} & \multirow{2}{*}{$> \mathrm{max} \left( 5,\frac{158}{\sqrt{P_{\mathrm{days}}}} \right)$} & \multirow{2}{*}{$<25$} & \multirow{2}{*}{$>\frac{20\;000}{P_{\mathrm{days}}}$}  & - \\
pseudo-circular &&&&&& \\
&&&&&& \\
\hline    
&&&&&& \\
VIMF                   &  7 & $\frac{D}{\sigma_D}$                 & $>20$ & $<25$ & $>30$        & -   \\
&&&&&& \\
\hline
\end{tabular}\\
*: {\footnotesize These selection conditions do not apply to orbital solutions from main processing when they have been confirmed by a SB1 spectrocopic solution, as explained in \cite{DR3-DPACP-178}.}
\label{tab:threshold}
\end{table*}


\section{The preprocessing}
\label{Sect:preprocessing}

\subsection{Rejection of outliers}
\label{subsect:outliers}

Each transit of a star through the {\it Gaia} astrometric instrument comes down to the measurement of two coordinates giving its position
relative to a reference point assigned to the star \citep[see][]{Lindegren21}. Of these, only the coordinate along the scan axis,
called hereafter the AL abscissa, $w$, is taken into account, being considerably more accurate than the other.
The AL abscissae were measured along the 9 CCD transits that constitute each field-of-view (FoV) transit. As the duration of the FoV transit is only a few seconds while we are looking for binary stars with periods of at least several days, the 9 CCD transits amount to 9 measurements of the same quantity. Outliers were therefore detected by intercomparison between the CCD transits. In practice, each AL abscissa was compared to the median of the 9 values, and the transit CCD was rejected when the difference exceeded 5 times the uncertainty of the abscissa.

This rejection process was not the only one that was applied, and subsequently transits were rejected after the calculation of some astrometric solutions. The largest residual transit was rejected when it exceeded 5 times the uncertainty, and the solution was recalculated; however, this operation was subject to three restrictions: (1) it was only done for linear models, i.e. the single star model and the two acceleration models, (2) the $\chi^2$ of the solution should exceed the number of transits used multiplied by 1.41 and (3) the proportion of rejected transits was limited to 5~\%.

\subsection{Correction of the perspective acceleration}
\label{subsect:perspective}

The distance from the Sun and the apparent position of a nearby star are varying during the
time-span of the {\it Gaia} mission, inducing slight changes in astrometric parameters which are
usually assumed to be constant. These so-called ``perspective effects'' were taken into account
in the calculation of the astrometric parameters of a few single stars in the Hipparcos Catalogue~\citep{hip}.
\citet{Dravins99} and \citet{Lindegren03} have shown that these effects are so closely related to radial 
velocity that they could be used to measure it. For {\it Gaia}, \citet{Halbwachs09} has shown that perspective effects 
do not significantly affect the astrometric orbits of binary stars; however, they do result in a ``perspective acceleration''
of stars that could be mistaken for an acceleration due to orbital motion. 

The perspective acceleration is based on the radial proper motion, $\mu_r$, which is defined as the radial velocity, $v_r$, divided
by distance. In practice, $\mu_r$, expressed in yr$^{-1}$, is related to $v_r$ and parallax, $\varpi$ by the following 
relationship:

\begin{equation}
\mu_r = v_r \; \varpi \times \frac{24 \; \pi \times \mathrm{year_{days}}}{180 \times \mathrm{au_{m}}}
\label{eq:mur}
\end{equation}

\noindent
where $v_r$ is in km.s$^{-1}$, $\varpi$ is in mas, $\mathrm{year_{days}}$ is the duration of the
year in days, and $\mathrm{au_{m}}$ is the length of the astronomical unit (au) in meters.
Perspective acceleration changes the AL abscissa of a star, $w$, by adding the following quantity:

\begin{equation}
\Delta w = - \mu_r \; (t-T) \times \; \left( \frac{\partial w}{\partial \varpi} \; \varpi + \frac{\partial w}{\partial{\mu_{\alpha *}}} \; \mu_{\alpha *} + \frac{\partial w}{\partial{\mu_\delta}} \; \mu_\delta \right)
\label{eq:Deltaw}
\end{equation}

\noindent
where $t$ is the epoch in years, and $T$ is an epoch close to the middle of the DR3 mission (in practice, $T=2016.0$). 
The partial derivatives of the abscissa with respect to the parallax and to the proper motion coordinates are those of the 5-parameter single star model.
If $\mu_r$ is unknown, then the abscissae are related to the parameters by a system 
of non-linear equations. Fortunately, the radial velocities of the bright stars are measured by {\it Gaia}, and a calculation based on the single star model, ignoring the perspective acceleration, already gives a good approximation of the parallax and of the proper motion. Thus, $\mu_r$
may be determined from Eq.~\ref{eq:mur}.

There are then two possibilities to take into account the perspective acceleration: either multiply the partial derivatives by $\mu_r (t-T)$, as they are in 
Eq.~\ref{eq:Deltaw}, or correct the abscissae by subtracting the perspective acceleration contribution, $\Delta w$. Because of its simplicity, we followed 
the latter method: $\Delta w$ was calculated from Eq.~\ref{eq:Deltaw}, using preliminary values of $\varpi$, $\mu_{\alpha *}$ and
$\mu_\delta$, and was subtracted to the AL abscissa. This correction was made for transits of all stars of known radial velocity when they were closer
than 200~pc, perspective effects being negligible beyond that. The astrometric binary solutions obtained for stars whose measurements have been corrected for perspective acceleration are identified by a flag in the solution tables.

\subsection{New calculation of the single star solution}
\label{subsect:singleStarSolution}

The above operations may have changed the transits sufficiently that the star should no longer be considered a potential binary.
For this reason, the single star solutions were recalculated, and they were accepted when $F_2$ was zero or negative. Of the $4\,115\,743$
stars that entered the processing chain, only 28 left at this stage. The rest were passed through the acceleration models.


\section{The acceleration solutions}
\label{sec:acc}

Acceleration models were already introduced in the reduction of the Hipparcos catalogue in order to describe binary stars that could not be covered by 
the orbital model because their period was too long. Two models were considered: the constant acceleration model, where the trajectory is a
parabola, and the variable acceleration model, which includes the time derivatives of the acceleration. 
The calculation of the solutions from these models was done with rejection of outliers, as explained at the end of Sect.~\ref{subsect:outliers}.
 

\subsection{The constant acceleration model}
\label{subsec:cstAcc}

Under the effect of the gravitational force, the two components of a binary are accelerated towards each other. If they
cannot be resolved and the photocentre does not coincide with the barycentre, the apparent motion of their photocentre is accelerated;
when the orbital period is much
longer than the time interval covered by the observations, the acceleration is nearly constant in intensity and direction.
These physical considerations are at the origin of the constant acceleration model.
Each coordinate of the photocentre varies with time according to a parabola.
However, fitting a parabola to the set of observations means setting the mid-mission position of the star as given by the first 
two terms of the solution, so that it is tangent to the parabola, and therefore far from the mean position of the star. To 
provide a position closer to the mean position of the star, we have followed the method explained in the Hipparcos 
catalogue \citep[vol. 1, Sect. 2.3.3]{hip}, and the partial
derivatives of the abscissa with respect to the acceleration, $(g_{\alpha *},g_\delta)$, are given by the following
equations:

\begin{equation}
\begin{array}{ll}
\frac{\partial w}{\partial g_{\alpha *}} & = \frac{1}{2} \frac{\partial w}{\partial \alpha *} \; 
\left[ (t-T)^2 - \frac{\Delta T^2}{3} \right] \\
\\
\frac{\partial w}{\partial g_\delta} & = \frac{1}{2} \frac{\partial w}{\partial \delta} \; 
\left[ (t-T)^2  - \frac{\Delta T^2}{3} \right] \\
\label{eq:derparg}
\end{array}
\end{equation}

\noindent
where $T$ is 2016.0 as in Sect.~\ref{subsect:perspective}, and $\Delta T$ is half 
the time interval covered by the observation of the star. For simplicity, the same value of $\Delta T$ was assumed for all stars, and we have adopted half
the time between the start of the Gaia science mission and the last DR3 observation, which, according to the information available when the software was finalised, lead to $\Delta T = 1035/2 = 517.5$~d, or 1.417~yr. This is an overestimation, however: in fact, the last observation of the DR3 was made on 28 May 2017, 3 days later than we had assumed,
but, on the other hand, the astrometric data of the first month of the Gaia scientific mission were not taken into account \citep{Lindegren21}. Moreover, the actual duration covered by the observations of a star is necessarily shorter than the duration covered by all the observations. For all these reasons, $\Delta T$
should be rather of the order of 470 days. However, the resulting shift in the mean position is very small: 0.06 mas times the acceleration in mas.yr$^{-2}$.

The significance of the constant acceleration model is defined from the acceleration vector. It is derived
from Eq.~\ref{eq:signif}, as explained in Sect.~\ref{subsubsec:signif}.


\subsection{The variable acceleration model}
\label{subsec:varAcc}

When the gravitational force is significantly changing over the duration of the mission, but when the
orbital period is still much longer than the observation time span, the time derivative of the acceleration,
$(\dot{g}_{\alpha *},\dot{g}_\delta)$, is added to the parameters of the model. However, the introduction of
these parameters could modify the proper motion of the star. In order to derive a proper motion corresponding to
the average displacement of the photocentre during the mission, 
the partial derivatives of the abscissa with respect to $\dot{g}_{\alpha *}$ and $\dot{g}_\delta$ are given 
by the equations:

\begin{equation}
\begin{array}{ll}
\frac{\partial w}{\partial \dot{g}_{\alpha *}} & = \frac{1}{6} \frac{\partial w}{\partial \alpha *} \; \left[ (t-T)^2 - \Delta T^2 \right] (t-T) \\
\\
\frac{\partial w}{\partial \dot{g}_\delta} & = \frac{1}{6} \frac{\partial w}{\partial \delta} \; 
\left[ (t-T)^2 - \Delta T^2 \right] (t-T) 
\label{eq:derpargp}
\end{array}
\end{equation}

\noindent
where $\Delta T$ is the same as in Sect.~\ref{subsec:cstAcc} above. We emphasise that the only effect of introducing $\Delta T$ is to produce a position and  proper motion similar to that which would result from applying the single star model. Since acceleration models cannot locate the barycentre of a binary, these parameters are still affected by the orbital motion of the photocentre relative to the barycentre.

The significance of the variable acceleration model is defined from the $(\dot{g}_{\alpha *},\dot{g}_\delta)$ vector. 
As above, it is derived from Eq.~\ref{eq:signif}, as explained in Sect.~\ref{subsubsec:signif}.

It is worth noticing that the acceleration models are both linear models, and that the solutions are found by solving a system of linear equations by singular value decomposition. 

In the Hipparcos reduction, the constant acceleration solution was considered as acceptable when the acceleration was significant, 
but the variable acceleration solution was always tried, and, when it was significant, it was preferred to the constant acceleration solution. We have 
kept this strategy, and, for that reason, the variable acceleration model was tried first.
The variable acceleration solution was
retained when it satisfied the conditions: $s > 12$ and $F_2 < 25$. Otherwise, the constant acceleration
solution was derived, and it was accepted when it satisfied the same conditions. Because of this acceptance, 
no other models were tried. Therefore acceleration solutions were retained for many binary stars with periods shorter than
the duration of the {\it Gaia} mission. This choice was originally made to save computing time;
this reason was no longer relevant at the time of processing the DR3, but the strategy remained unchanged
because it was considered that the risk of producing an orbit with a wrong period was more serious than 
that of giving only a polynomial fit for a trajectory that could be described by an ellipse.
This is a matter of judgement which may be considered questionable, and will probably be approached differently in
the next data release.


\subsection{Selection of the acceleration solutions}
\label{subsect:selAcc}

To delimit a domain of acceptance of the solutions, we consider, for both acceleration models, the true acceleration
projected on the sky, called $\Gamma$ in the following. $\Gamma$ is derived from the equation:

\begin{equation}
\Gamma = \frac{\sqrt{g_{\alpha *}^2 + g_{\delta}^2}}{\varpi}
\label{eq:Gamma}
\end{equation}

When the apparent acceleration is expressed in mas.yr$^{-2}$ and when the parallax is
in mas, $\Gamma$ is obtained in au.yr$^{-2}$.

In order to estimate a maximum value for $\Gamma$, we consider a bright star with a dark companion, so that the acceleration
of the photocentre is that of the bright star. Applying the law of gravitation, and assuming that the distance between the
components is approximately equal to the semi-major axis of the orbit of the bright star around the companion, we deduce from Kepler's third law that the acceleration is then:

\begin{equation}
\Gamma_{\mathrm{au/yr^2}} \approx 4 \pi^2  \left( \frac{{\cal{M}_{\cal{M}_\odot}}}{P_\mathrm{yr}^4} \frac{q^3}{(1+q)^2}\right)^{1/3} 
\label{eq:gammaG}
\end{equation}

\noindent
where $P$ is the period, $\cal{M}$ the mass of the bright star and $q$ the mass ratio, i.e. the ratio between the mass
of the dark companion and $\cal{M}$.

\begin{figure*}
\begin{subfigure}{.5\textwidth}
  \centering
  \includegraphics[width=0.9\linewidth]{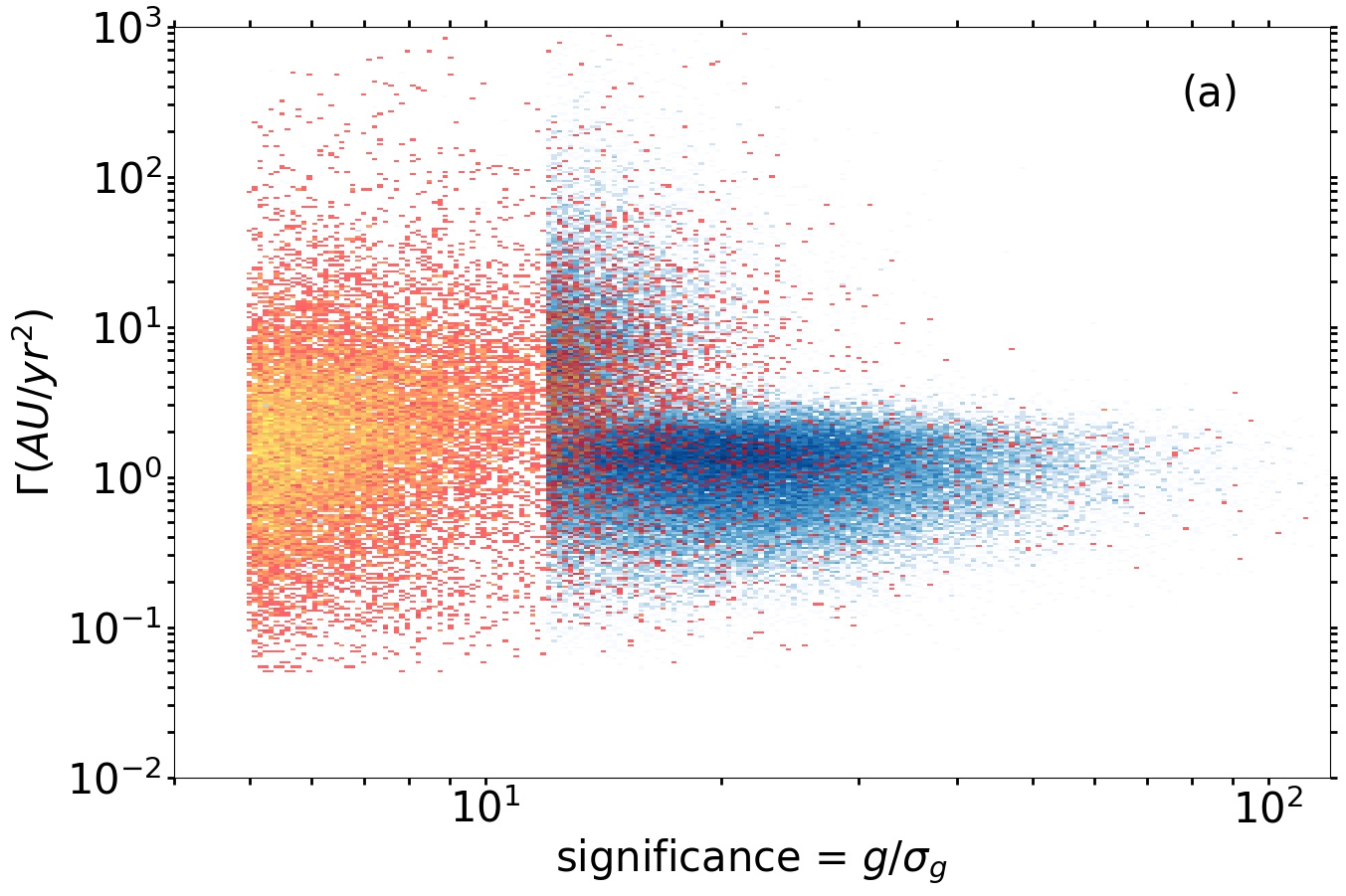}
\end{subfigure}
\begin{subfigure}{.5\textwidth}
  \centering
    \includegraphics[width=0.9\linewidth]{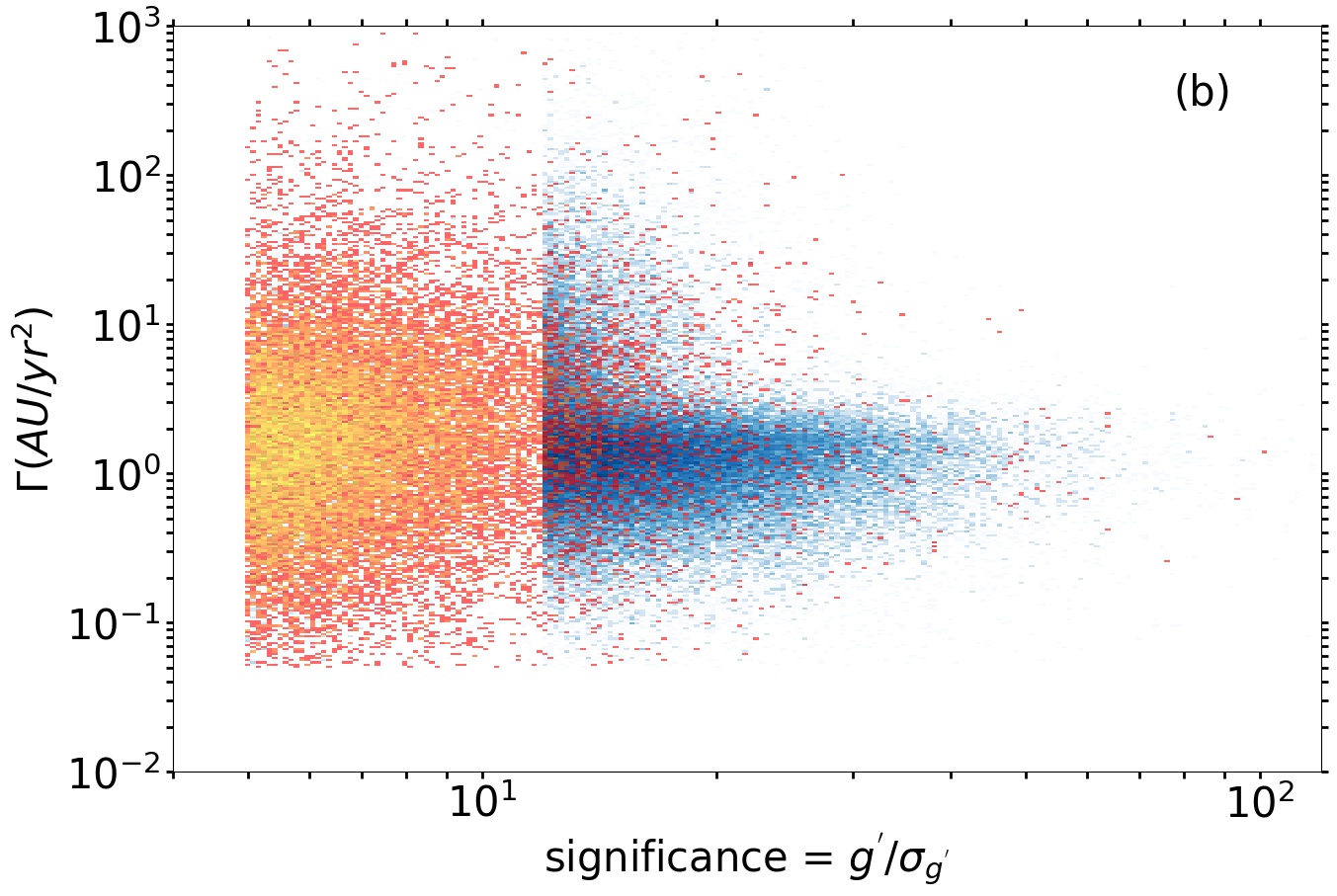}
\end{subfigure}
\begin{subfigure}{.5\textwidth}
  \centering
  \includegraphics[width=0.9\linewidth]{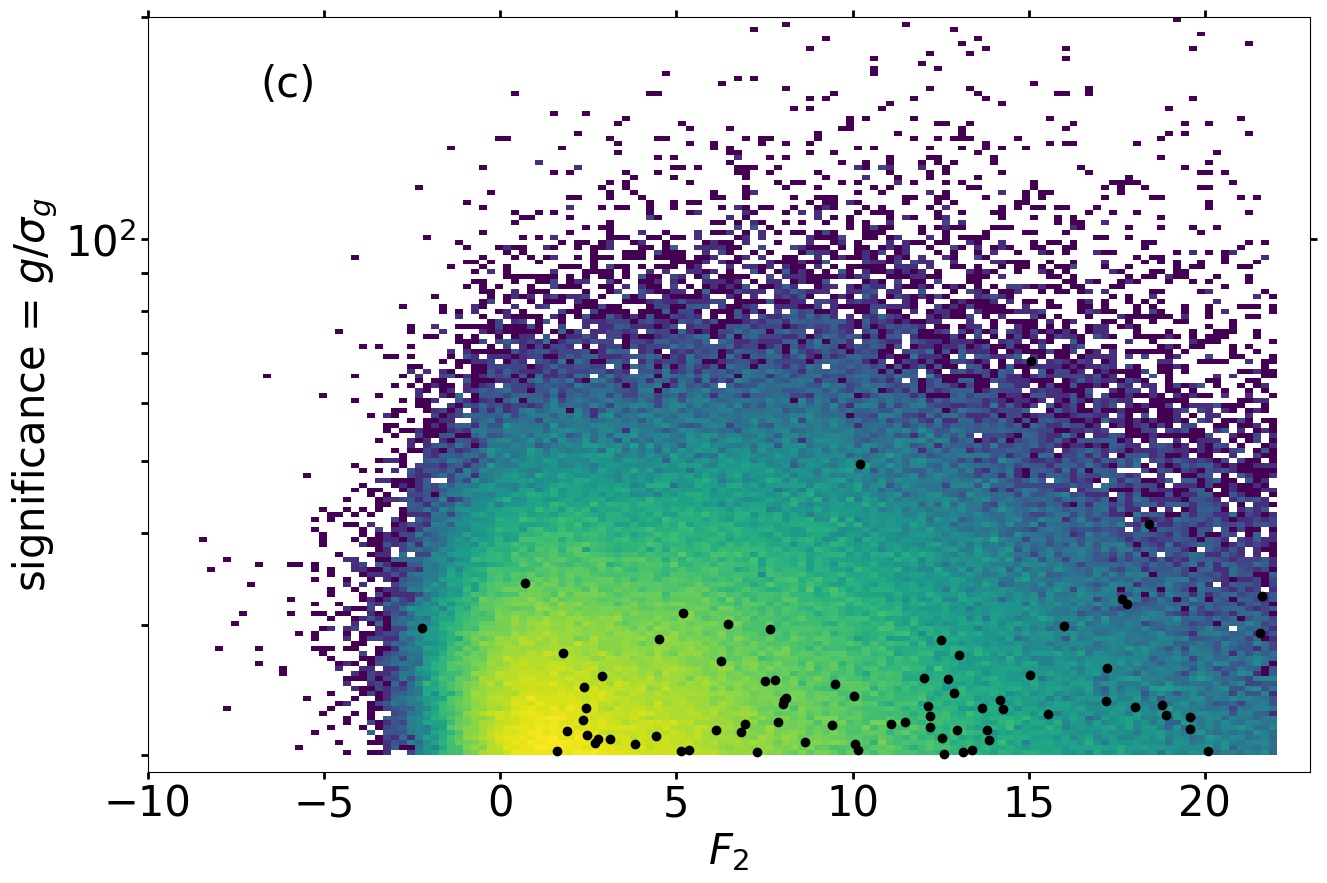}
\end{subfigure}
\begin{subfigure}{.5\textwidth}
  \centering
    \includegraphics[width=0.9\linewidth]{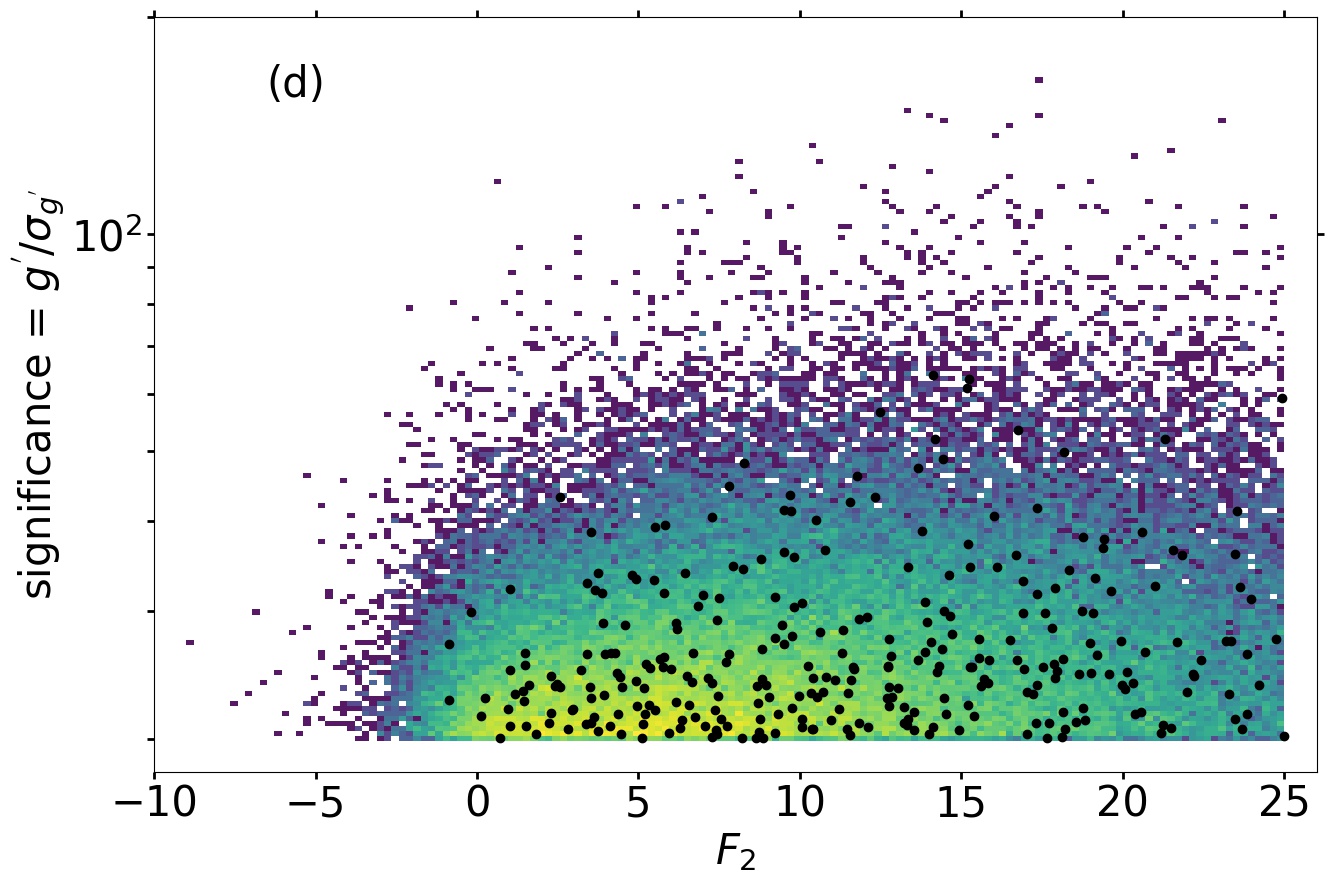}
\end{subfigure}
\caption{Selection of the acceleration solutions. The left-hand panels correspond to solutions with constant acceleration, and those on the right to those with variable acceleration. 
Panels (a) and (b): Significance vs. acceleration diagrams from a processing trial without the selection based on parallax error. 
The solutions with $F_2$ below the final threshold are in blue, while those above are in orange.
The solutions of $5<s<12$ were accepted in the absence of better solutions from another model, while the others were immediately
accepted when $s>12$ and $F_2$ below 22 (for constant acceleration solutions) or 25 (for variable acceleration solutions).
The amount of solutions with a mean acceleration $\Gamma > 3 \; \mathrm{au/yr^2}$ falls off when the significance exceeds 20. Panels (c) and (d): 
the $F_2$ vs. $s$
diagrams of the solutions finally selected in the post-processing. The black dots are the remaining solutions with $\Gamma > 3 \; \mathrm{au/yr^2}$; they still correspond to the
smallest significances, but this essentially due to the filtering based on the parallax error. 
  }
\label{fig:acc}
\end{figure*}

An upper limit of $\Gamma$ is obtained when the bright star is a giant solar-mass star with a companion still on the
main sequence or on the subgiant branch, with $q$ slightly less than 1, and when the period is twice the duration of the
{\it Gaia} mission. Such a system is not frequent, but not exceptional in a sample bounded by a limit in magnitude.
With a period of 5.7~yr, the acceleration may be as large as about 2.4~$\mathrm{au/yr^2}$ for a circular orbit; over half a period, the average acceleration can then be as high as 2.0~$\mathrm{au/yr^2}$. This result is however a rough 
estimate: We have neglected the eccentricity, which will produce a larger acceleration near
the periastron, and we have assumed a period much longer than the duration of the mission, whereas we know that many solutions
are for shorter periods. 
Considering the distribution of values obtained from the solutions (see Fig.~\ref{fig:acc}, (a) and (b)), 
we see a concentration below $\Gamma=3 \; \mathrm{au/yr^2}$, which we adopt as the limit for an acceptable acceleration solution.
Stars whose acceleration is really beyond this limit are certainly very rare, but they are also objects of great scientific interest,
such as binaries with a massive dark component, and for this reason we do not reject them directly. 

The following conditions were introduced to restrict the proportion of $\Gamma$ larger than this limit:

\begin{enumerate}

\item As mentioned in Sect.~\ref{subsec:varAcc} above, the main processing was performed with the threshold 
$s > 12$ for direct acceptance and $s > 5$ for the alternative solutions. In addition, conditions based on 
the significance of the parallax were introduced for both levels:
$\varpi/\sigma_\varpi \; > \; 1.2 \; s^{1.05}$ for the
constant acceleration model, and  $\varpi/\sigma_\varpi \; > \; 2.1 \; s^{1.05}$ for the variable acceleration model.
These conditions have been set because they very effectively reduce the amount of high accelerations; it is worth noticing that, in fact,
they are not criteria based on the quality of the solutions: reading them from right to left, they both mean that,
for a given distance, the most significant solutions are rejected. This results in rejecting the largest accelerations because many of them are highly
significant, and their occurence rate is thus artificially reduced. However, it was necessary to apply these criteria to avoid too much pollution by partially resolved double stars. Another beneficial consequence of this filter is that the short period binaries whose acceleration solution was rejected in this way were able to have an orbital solution.
Including alternative solutions, the main processing yielded $808\,992$ constant acceleration solutions and $569\,022$ variable
acceleration solutions. These solutions must be filtered in the post-processing to keep only those that seem most likely to be real.

\item The significance, $s$, must be over 20 for both acceleration models. 

\item The goodness of fit, $F_2$, must be below 22 for the constant acceleration model, and below 25 for the variable acceleration model. 
The limit is more severe for the constant acceleration solutions so that the sequences of the HR diagrams deduced from the parallaxes of these solutions are always thinner than those deduced from the single star model. 

\end{enumerate}

Conditions (2) and (3) reduced the number of solutions to $246\,798$ for constant accelerations, and $91\,227$ for variable accelerations.

The two lower panels of Fig.~\ref{fig:acc} show the distributions of the selected solutions in the $F_2$ vs.
$s$ diagrams. Thanks to the parallax filtering, item (1) above, the solutions with $\Gamma>3$~au.yr$^{-2}$ are very rare: 0.03~\% 
of the constant acceleration solutions, and 0.3~\%
for the variable accelerations.
If filters (2) and (3) are applied to a test sample unaffected by filter (1), the proportions of large $\Gamma$ are 3.1 and 5.6~\% respectively.
Unless one is prepared to sacrifice the vast majority of solutions, these high rates resist the quality criteria we have tried; this issue is still
being discussed in Sect.~\ref{sect:discussion}, in the light of independent investigations. Several origins have been considered for the
high acceleration solutions: short-period binaries, partially resolved binaries or outlier astrometric measurements. We can only conclude that the
stars with acceleration solutions
are most likely genuine binaries, but the physical interpretation of the accelerations is hazardous.


\section{The orbital solutions}

An orbital solution is calculated when the orbit has been observed either in its entirety or over a sufficient portion to extrapolate the missing part.

\subsection{Calculation of the orbital solutions}

The abscissa of an unresolved astrometric binary can be written as in the following equation:

\begin{equation}
\begin{array}{ll}
w = & w_B + \frac{\partial w_B}{\partial \delta} (\cos E - e) \times A + \frac{\partial w_B}{\partial \alpha*} (\cos E - e) \times B \\
 & + \frac{\partial w_B}{\partial \delta} \sqrt{1-e^2} \sin E \times F  + \frac{\partial w_B}{\partial \alpha*} \sqrt{1-e^2} \sin E \times G
\end{array}
\label{eq:wAO}
\end{equation}

\noindent
where $w_B$ is the abscissa of the barycentre, which is given by the single star model, $A$, $B$, $F$ and $G$ are the Thiele-Innes (TI, hereafter) elements,
whose definition is given in Eq.~\ref{eq:ABFG}, $e$ is the eccentricity of the orbit, and $E$ is the eccentric anomaly. $E$ depends on the
observation epoch, but also on the orbital period, $P$, on $e$, and on the periastron epoch, $T_0$, so the model is finally based on 12 unknowns:
the 5 parameters of the single star solution, and $A$, $B$, $F$, $G$, $P$, $e$, $T_0$; the partial derivatives of $w$ with respect to the TI
elements are given in Eq.~\ref{eq:wAO}, and those with respect to $P$, $e$ and $T_0$ can be found in \citet{Goldin06}.

The derivation of the orbital solutions is based on the Levenberg-Marquardt algorithm, but this can only lead to the solution if it starts from a good starting point.
In practice, a calculation starting from $e=0$ often leads to the right result if the period tested is close to the actual period, but scanning a grid covering the space of $P$, $e$ and $T_0$ is still 
necessary to be sure to find the solution. The period range tried is from 10 days to the duration covered by the observations of the star divided by 0.6. The solutions are calculated and published in terms of TI elements, but the calculation of the elements usually used to describe an orbit, which are the so-called Campbell elements (semi-major axis, inclination of the orbit, position angle of the ascending node and periastron longitude), is presented in Appendix~\ref{Sect.ABFG2Campbell}, where the calculations of their uncertainties are also given. 
The extension of this calculation to the Thiele-Innes elements introduced by both astrometric and spectroscopic orbits is presented in Appendix~\ref{Sect.CH2Campbell}.

The significance of the orbital model is defined as the semi-major axis of the orbit, $a_0$, divided by its uncertainty.
This definition of the significance was adopted since a significant orbit is expected to have a large semi-major axis, when, on the contrary, 
an orbit with a semi-major axis close to or smaller than its uncertainty may be obtained for a single star. The semi-major
axis is not explicitly a parameter of the orbital model that we have used, but it may be derived from the TI
elements, thanks to Eq.~\ref{eq:a}. Its uncertainty is derived from Eq.~\ref{eq:tABFG} and Eq.~\ref{eq:siga}.

\subsection{Pseudo-circular and circular orbital solutions}

Rare orbits (8 in total) have been directly calculated as circular and are given with $e=0$. These calculations required special arrangements, which have been also adopted for orbits with very small eccentricities, as explained below.

Preliminary calculations showed that the eccentricities below 0.006 were all at least three times smaller than their uncertainties. For such solutions, the uncertainties of the TI elements are also often very large, as is the uncertainty of $T_0$, which can be larger than the period. These anomalies arise from the indeterminacy of the position of the periastron, affecting the TI elements through their dependence on the longitude of the periastron, $\omega$ (see Eq.~\ref{eq:ABFG}). They do not affect the validity of the solution, for which the Campbell elements calculated as explained in Appendix~\ref{Sect.ABFG2Campbell} have correct uncertainties, except for $\omega$, whose uncertainty far exceeds $2\pi$; however, they may confuse the user. 

In order to limit these effects, orbits with eccentricity less than 0.0005 have been ``pseudo-circularised'' in the following way: 
The Campbell elements were derived, and the periastron longitude, $\omega$, was set to 0 in order to place the periastron on the ascending node. Therefore, the periastron epoch, $T_0$, has been advanced by
subtracting the time interval corresponding to the orbit section between the ascending node and the initial periastron, which for a circular orbit is $P \times \omega/(2\pi)$. As the initial orbit is not circular, this formula is an approximation that
is only possible for very small eccentricities;
this is why it was only done for $e<0.0005$. The TI elements were then recalculated, keeping the semi-major axis, orbit inclination and line-of-nodes position angle of the initial solution. 

After these transformations, the solution is based on only 10 unknowns: $e$ and $T_0$ are fixed, and only 3 of the 4 TI elements remain, since $\omega=0$ implies that $G=-AF/B$, or $F=-BG/A$. Therefore, $G$ was no longer considered as an unknown provided that $|B|>10^{-5}$~mas. Otherwise, it was $F$ that had to be removed, but this did not happen in practice. Thus, the solution variance-covariance matrix was recalculated for the 10 selected unknowns. This $10 \times 10$ matrix was inserted into the $12 \times 12$ matrix of orbital solutions, setting the unnecessary terms to 0. 

In the end, this gives quite reasonable uncertainties for the TI elements and for $T_0$. However, the solution will give an uncertainty and correlation terms that are zero for $G$, which is wrong and needs to be corrected in order to deduce the uncertainties of Campbell's elements as explained in Appendix~\ref{Sect.ABFG2Campbell}. The uncertainty of $G$, $\sigma_G$, is derived from the equation:

\begin{equation}
\begin{array}{rl}
\sigma_G = |G| & \left[\left(\frac{\sigma_A}{A}\right)^2 + \left(\frac{\sigma_B}{B}\right)^2 + \left(\frac{\sigma_F}{F}\right)^2 \right.\\ 
           & \left. + 2\left(\frac{\rho_{AB}\sigma_A\sigma_B}{AB} + \frac{\rho_{AF}\sigma_A\sigma_F}{AF}  + \frac{\rho_{BF}\sigma_B\sigma_F}{BF}\right)\right]^{1/2}
\end{array}
\label{eq:cu4nss_astrobin_orbital_sigG}
\end{equation}

The correlation coefficients relating $G$ to the other TI elements are:

\begin{equation}
\begin{array}{rl}
\rho_{AG} = & +1 \\
\rho_{BG} = & -1 \\
\rho_{FG} = & +1
\end{array}
\label{eq:cu4nss_astrobin_orbital_rhoG}
\end{equation}

These modifications must also be made to circular orbits if the uncertainties of their Campbell elements are to be calculated.

\subsection{Selection of the orbital solutions}

\begin{figure*}
\begin{subfigure}{.5\textwidth}
  \centering
  \includegraphics[width=0.9\linewidth]{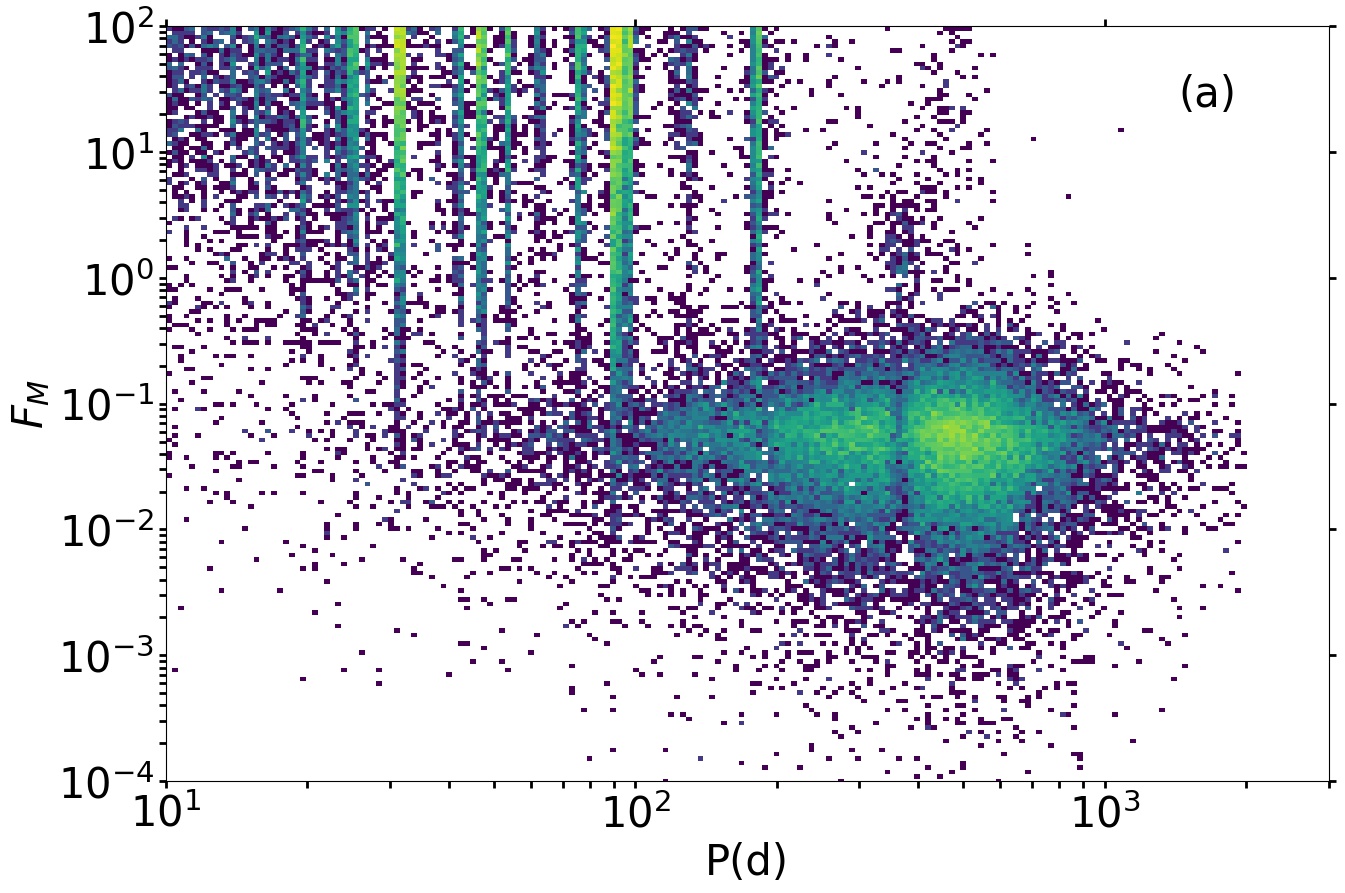}
\end{subfigure}
\begin{subfigure}{.5\textwidth}
  \centering
    \includegraphics[width=0.9\linewidth]{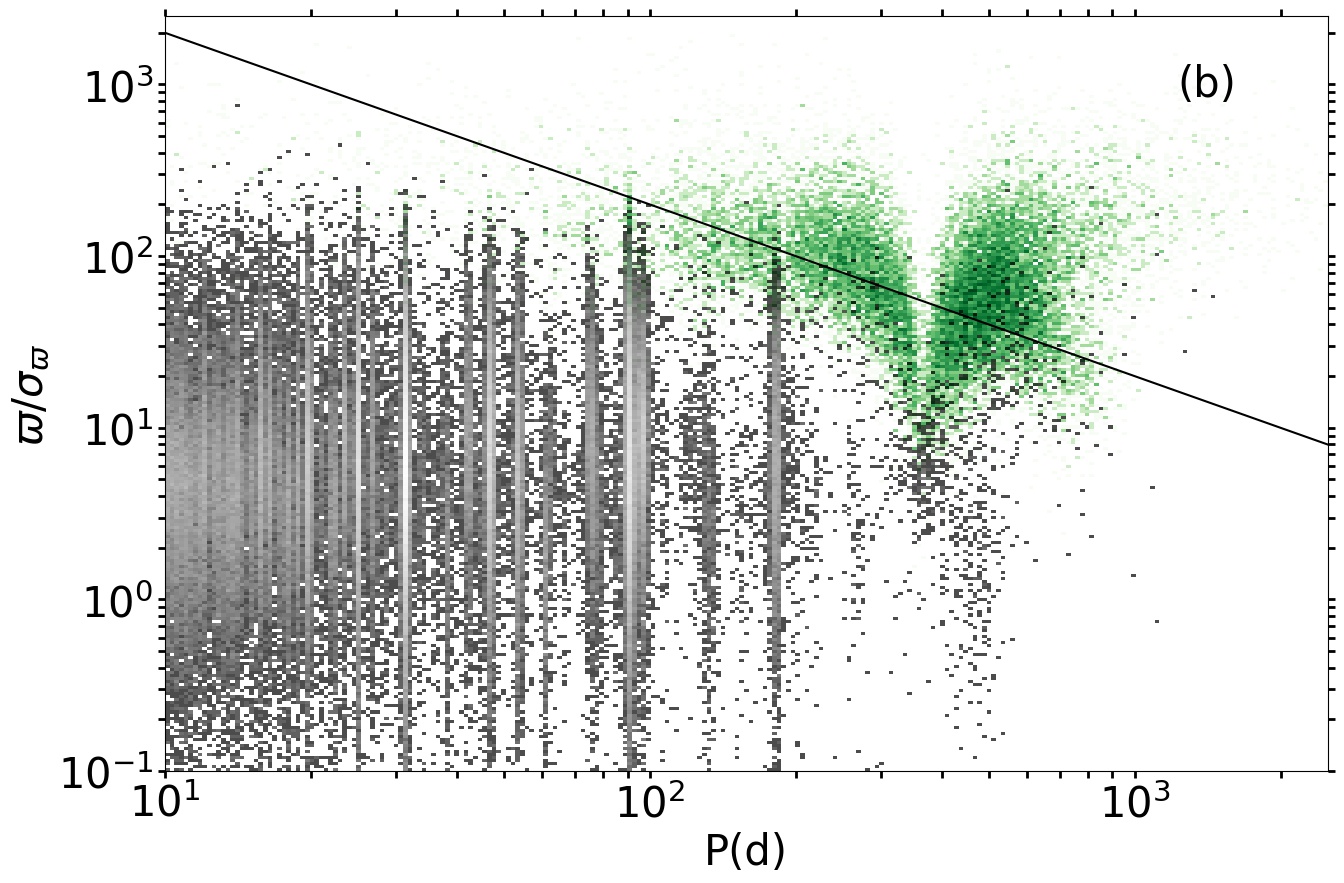}
\end{subfigure}
\begin{subfigure}{.5\textwidth}
  \centering
  \includegraphics[width=0.9\linewidth]{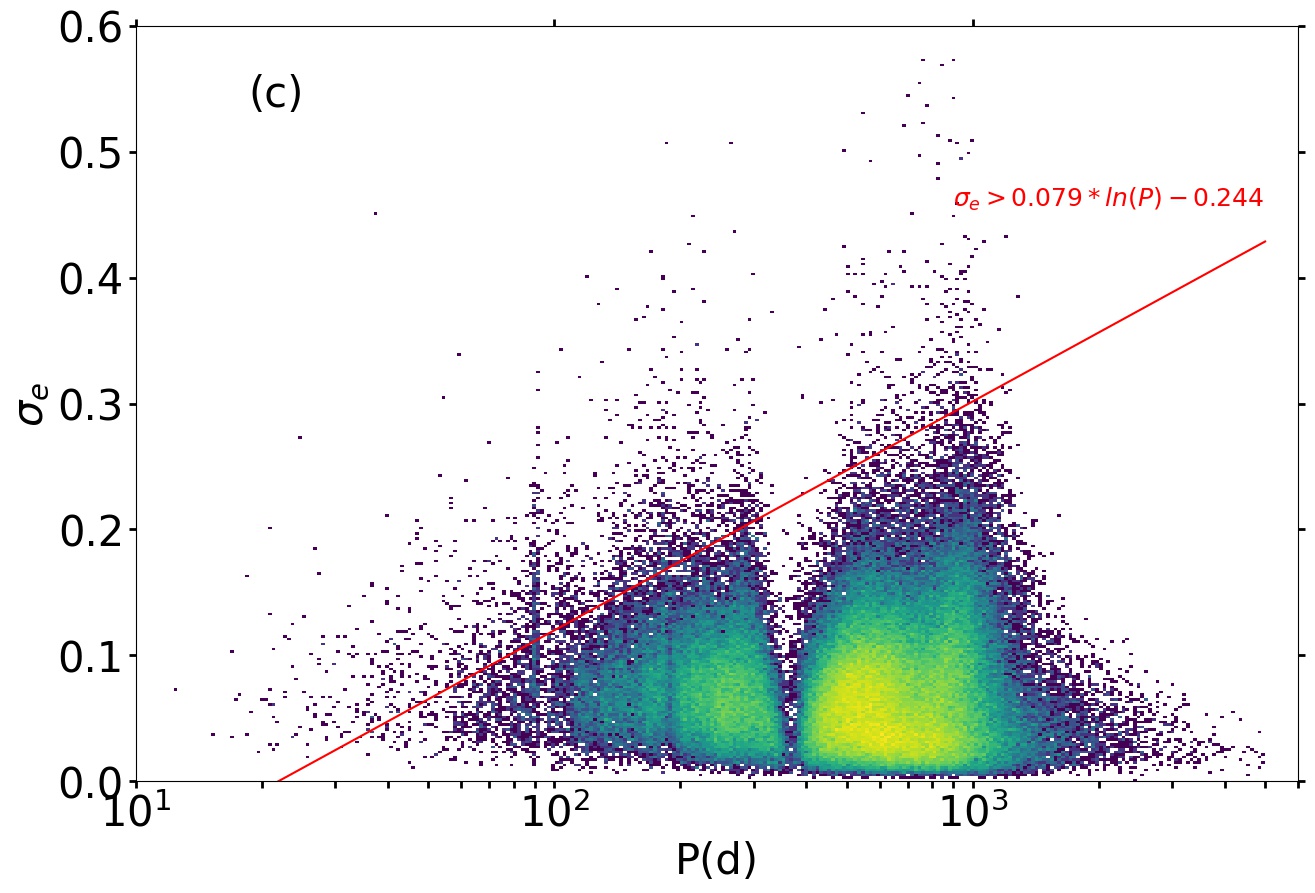}
\end{subfigure}
\begin{subfigure}{.5\textwidth}
  \centering
    \includegraphics[width=0.9\linewidth]{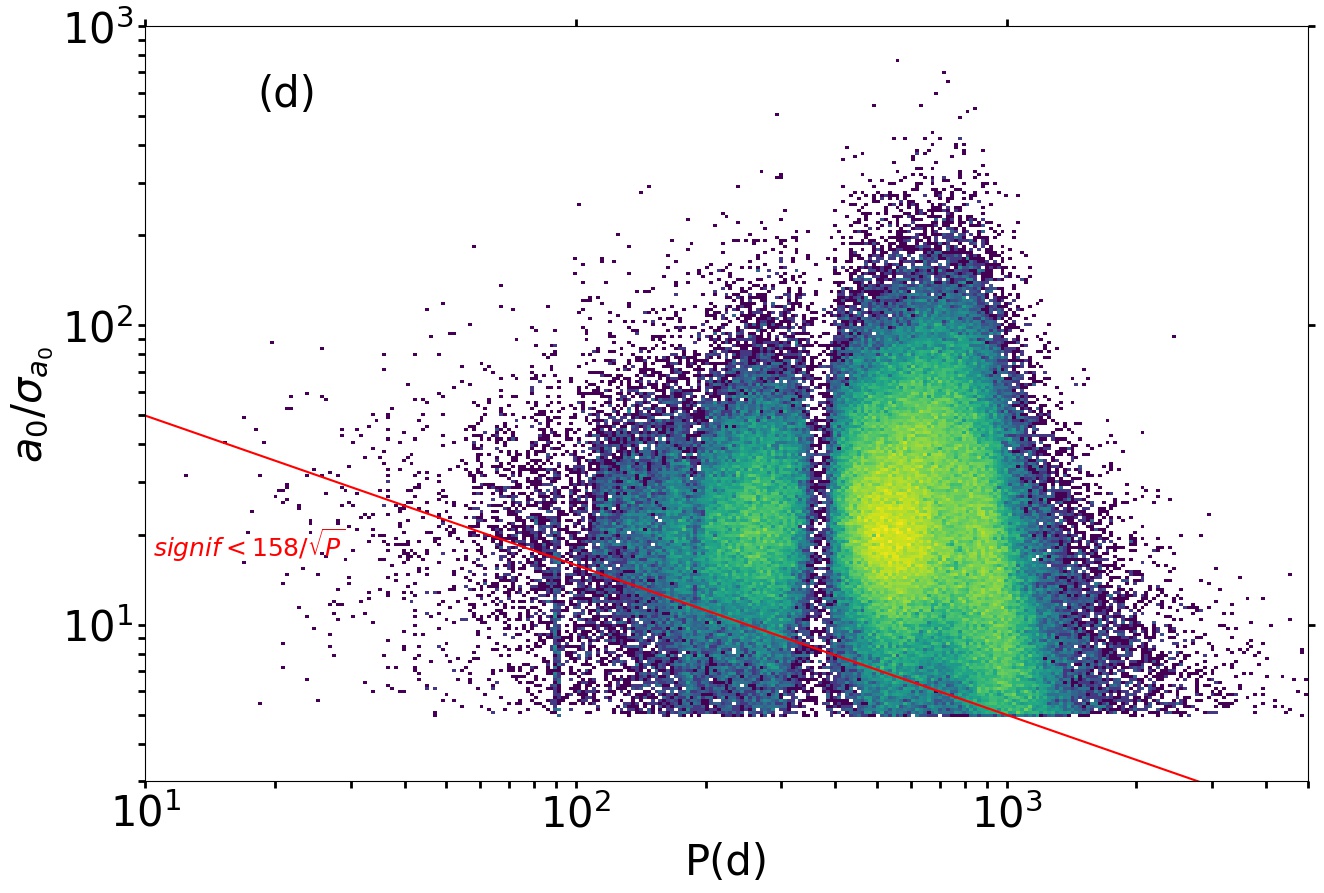}
\end{subfigure}
\begin{subfigure}{.5\textwidth}
  \centering
    \includegraphics[width=0.9\linewidth]{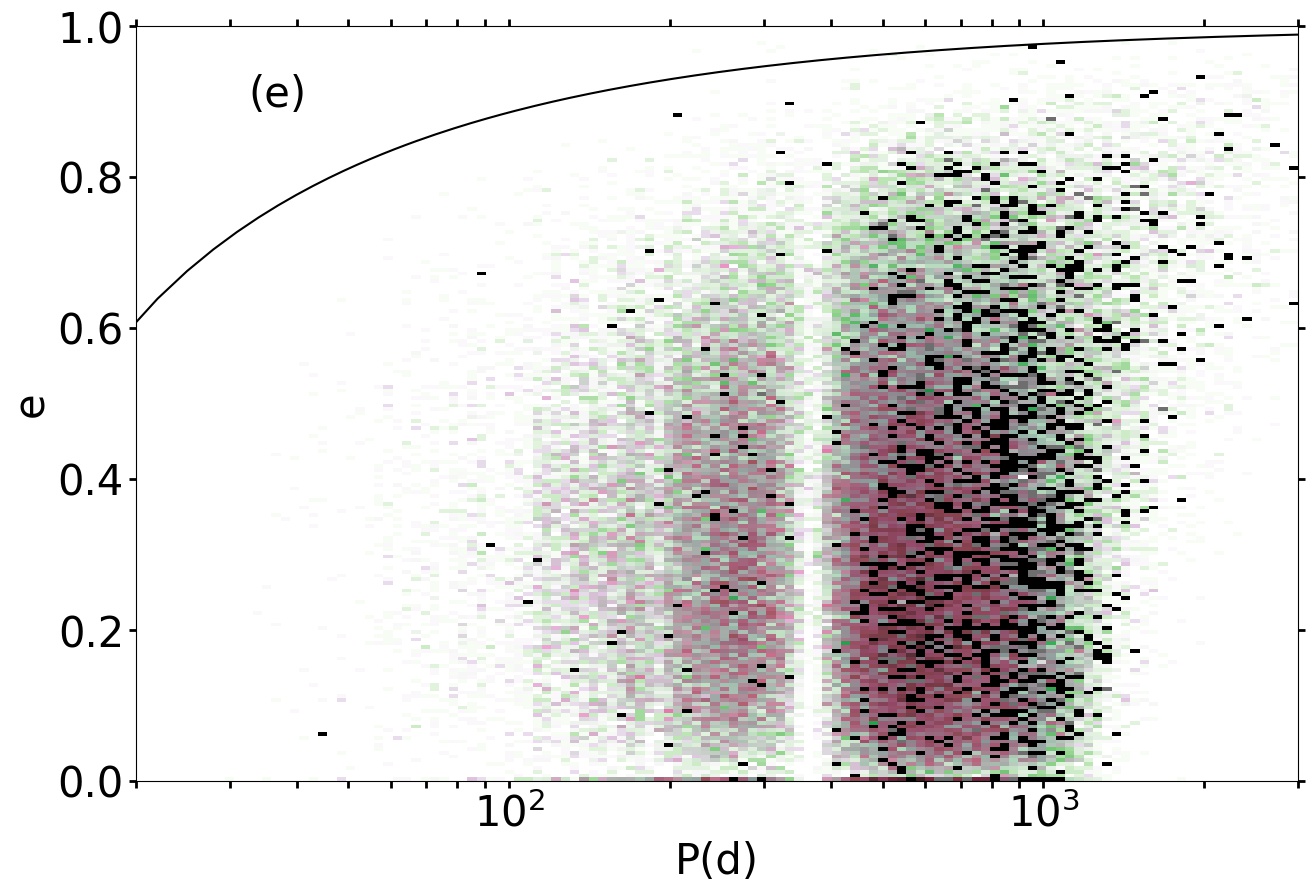}
\end{subfigure}
\begin{subfigure}{.5\textwidth}
  \centering
    \includegraphics[width=0.95\linewidth]{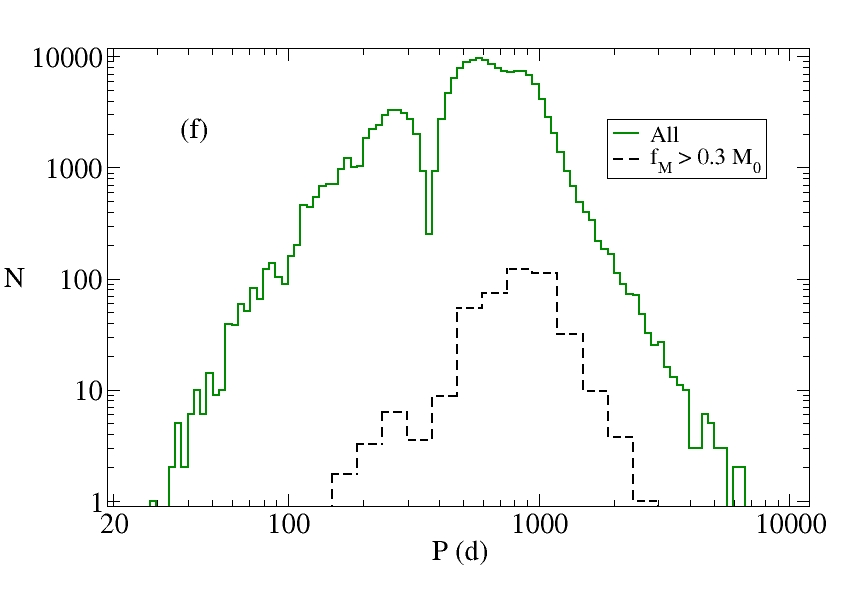}
\end{subfigure}
\caption{Selection of the astrometric orbital solutions. The first row refers to the solution obtained from a processing trial without the 
selection based on parallax error. (a): Period vs.
mass function diagram, showing a clump of plausible solutions, bounded by $f_{\cal M} \approx 0.3 \; {\cal M}_\sun$, and various sequences, organised around 
particular periods, which cross the clump and extend well beyond. (b): Period vs. ``parallax significance''. The solutions
with $f_{\cal M} < 0.3 \; {\cal M}_\sun$ are in green, and the large $f_{\cal M}$, which are well clustered around the same periods as in panel (a), are in black. The line is the
limit: $\varpi/\sigma_\varpi \; = \; 20\;000/P_d$ which was implemented in the main processing to discard the dubious solutions.
The second row shows the density map diagrams used to set up the post-processing filters to remove the concentrations still existing over specific periods. (c): Period--eccentricity uncertainty diagram; solutions above the red line were rejected (d): Period--significance diagram; solutions below the red line were rejected.
The third row shows properties of the DR3 solutions eventually selected, after applying the post-processing filters. (e): the period--eccentricity
diagram; the black curve is the maximum eccentricity assuming that the periods shorter than 10 days are circularized. The black dots are the solutions with $f_{\cal M} > 0.3 \; {\cal M}_\sun$. The astrometric-only orbital solutions are in green, while the solutions confirmed by a spectroscopic SB1 orbit from {\it Gaia} are in purple.
(f): histogram of periods; 
the proportion of solutions with mass function above 0.3 solar masses, in black, increases with period, as the gap between these solutions and the totality, in green, is reduced (in order to highlight possible overdensities, the step size of the green histogram is very fine; to avoid fluctuations in counts from small numbers, the black histogram was plotted by merging the count intervals into clusters of 4, and giving the mean value for each cluster). 
  }
\label{fig:AO}
\end{figure*}

The orbital solutions were first selected according to the thesholds set out in Sect.~\ref{subsubsec:signif}, the significance being calculated as explained in the previous section. In order to identify
doubtful solutions, we have taken into account the mass function, which is derived from:

\begin{equation}
f_{\cal M} =  \frac{{a_0}^3 \times 365.25^2}{P^2 \varpi^3}
\label{eq:fM}
\end{equation}

\noindent
where $P$ is the period in days, and where $a_0$ and $\varpi$ are in the same units; $f_{\cal M}$ is thus obtained in solar masses. 
Taking into account that $a_0$ refers to the orbit of the photocentre, the third Kepler's law gives the following expression:

\begin{equation}
f_{\cal M} =  \frac{\| {\cal F}_1 {\cal M}_2 - {\cal F}_2 {\cal M}_1 \|^3}{({\cal F}_1 + {\cal F}_2)^3 ({\cal M}_1+{\cal M}_2)^2} 
\label{eq:fMgene}
\end{equation}

\noindent
where ${\cal F}_1$ and ${\cal F}_2$ are the photometric fluxes of the brightest and of the faintest component, respectively.
${\cal M}$ refers to the masses. When ${\cal F}_2$ is negligible compared to ${\cal F}_1$, this equation becomes:

\begin{equation}
f_{\cal M} ({\cal F}_2 \ll {\cal F}_1)   = \frac{{\cal M}_1 q^3}{(1+q)^2}
\label{eq:fMdark}
\end{equation}

\noindent
where $q={\cal M}_2/{\cal M}_1$ and where the index ``1'' refers to the brightest component.

Panel (a) of Fig.~\ref{fig:AO} shows the period vs. mass function diagram. There are two kinds of false solutions:

\begin{itemize}
\item Solutions of $f_{\cal M} > 0.3 {\cal M}_\sun$. This limit looks rather large for dwarf stars, but roughly acceptable for
binaries with a giant component and a component still close to the main sequence, or for a triple system with a massive pair
as a secondary component.
\item Solutions with particular periods, which probably depend on the 63-day precession period of the satellite. These solutions have
usually very large mass function,but they descend sufficiently into the small $f_{\cal M}$ range to contaminate plausible solutions.
\end{itemize}

A large part of the last category of false solutions was discarded by using a filter based on the parallax significance, as 
indicated in Fig.~\ref{fig:AO}b. This filter consists of selecting only the solutions that satisfy the following condition:

\begin{equation}
\varpi/\sigma_\varpi \; = \; 20\;000/P_d
\label{eq:signivarpiAO}
\end{equation} 

\noindent
and it allowed to keep $179\,234$ astrometric orbits as a result of the main processing.

Although the filter of Eq.~\ref{eq:signivarpiAO} has rejected most of the large mass function solutions, a few concentrations around particular periods still persist in the processing results (Fig.~\ref{fig:AO}c and d).
Most of these remaining false solutions were eliminated in the post-processing, where only solutions satisfying the following 
two filters were selected:

\begin{itemize}
\item A filter based on the quality of the eccentricity: $\sigma_e  < 0.079 \ln P - 0.244$, where $P$ is in days.
\item A filter based on the significance: $s \; = \; a_0/\sigma_{a_0} > 158/\sqrt{P}$, where $P$ is again in days.
\end{itemize}

After these operations, about $166\,500$ orbits were retained, and the concentrations along particular periods have all disappeared, as shown in Fig.~\ref{fig:AO}e.
Moreover, very few orbits have eccentricities beyond the 10-day circularisation limit, calculated according to \citet{Halbwachs05}.
The proportion of $f_{\cal M} > 0.3 {\cal M}_\sun$ is now
about 1~\%, but it can be seen in Fig.~\ref{fig:AO}f that it increases considerably with the period: for example, it is only 0.2~\% for periods of less than 1 year, but reaches 3.3~\% for periods longer than 1000 days.

Subsequently, all the main processing astrometric orbits were compared to the SB1 spectroscopic orbits derived from {\it Gaia} radial velocities \citep{DR3-DPACP-178}.
The matching of the orbits of the two types resulted in nearly $30\,950$ combined orbits, of which about 2470 were 
recovered after having been rejected by the post-processing. At the same time, almost 3500 post-processing orbital solutions were
rejected by the validation \citep{Babusiaux}, or discarded as redundant with solutions obtained with Markov Chain Monte Carlo and Genetic Algorithms \citep{Holl}. In the end, our contribution to the DR3 astrometric orbital solutions amounts to nearly 165500 orbits.


\section{The variability-induced movers}
\label{sec:VIM}

The variability-induced movers (VIM hereafter) were first described by \cite{Wielen}. They consist in unresolved binary
systems containing one photometrically variable component. The result is a photocentre which is moving between the 
components in accordance with fluctuations in the total brightness of the system. Simultaneously, it is following
the orbital motion, if this is perceptible over the duration of the mission. Several types of VIM were expected from
the {\it Gaia} astrometric and photometric measurements: when the orbital motion is not perceptible, the system is a fixed
VIM, or VIMF. When the orbital motion looks like a motion of the components relative to each other, in a straight line
and at constant velocity, the system is a VIM with linear motion, or VIML. This is followed by VIM with acceleration (constant or 
variable) or VIMA, and then VIM with orbital motion, or VIMO. Only VIMFs are described in detail below, as the others
could not be included in the DR3.


\subsection{The VIM with fixed components (VIMF) model}
\label{subsec:VIMFmodel}

The additional parameters specific to the VIMF model are the coordinates of the photocentre, 
$(D_{\alpha *}, D_\delta)$, measured from the variable component when the total photometric flux of the system
is equal to a reference flux, noted $\bar{\cal F}$. In practice, $\bar{\cal F}$ is the median flux of all transits of
the system. The partial derivatives of the abscissa with respect to $D_{\alpha *}$ an $D_\delta$ are given by
the following equations:

\begin{equation}
\begin{array}{ll}
\frac{\partial w}{\partial D_{\alpha *}} & =  \frac{\partial w}{\partial \alpha *} \; \left( \frac{\bar{\cal F}}{\cal F} - 1 \right) \\
\\
\frac{\partial w}{\partial D_\delta} & = \frac{\partial w}{\partial \delta} \; \left( \frac{\bar{\cal F}}{\cal F} - 1 \right) 
\end{array}
\label{eq:derparD}
\end{equation}

\noindent
where ${\cal F}$ is the photometric flux of the transit. Therefore, the position of the system and its proper motion also apply to the photocentre when the total flux is $\bar{\cal F}$.

The parameters of the VIMF model are thus derived from a linear system that is solved by singular value decomposition,  but the uncertainty of 
each transit deserves
a dedicated calculation, since it depends on the astrometric uncertainty of the measured abscissa, $\sigma_w$, but
also on the uncertainty of the photometric flux, $\sigma_{\cal F}$, on which an error from the model depends. For each 
transit, this error, $\sigma_{\mathrm{mod}}$, is calculated from the equation: 

\begin{equation}
\sigma_{\mathrm{mod}} = \sigma_{\cal F} \; \frac{\bar{\cal F}}{{\cal F}^2} \; \left| \left( \frac{\partial w}{\partial \alpha *} D_{\alpha *} + \frac{\partial w}{\partial \delta} D_\delta \right)  \right|
\label{eq:s_mod_VIMF}
\end{equation}

The total uncertainty of transit is then given by the equation:

\begin{equation}
\sigma_{\mathrm{VIMF}} = \sqrt{\sigma_w^2 + \sigma_{\mathrm{mod}}^2}
\label{eq:s_VIMF}
\end{equation}

The $\sigma_{\mathrm{VIMF}}$ uncertainties are used
to derive any VIMF solution, as well as $F_2$ and the correction coefficient $c$ as explained in Sect.~\ref{susubsec:F2}.
As $\sigma_{\mathrm{VIMF}}$ depends on $D_{\alpha *}$ and $D_\delta$, and thus on the VIMF solution, the calculation is 
iterative and the VIMF model cannot be considered linear in the strict sense.
The significance of the solution is defined from the modulus of the $(D_{\alpha *}, D_\delta)$ vector, as explained in Sect.~\ref{subsubsec:signif}.

As far as the VIML model is concerned hereafter, suffice it to say that it has two more parameters, which are 
the time derivatives of $D_{\alpha *}$ and $D_\delta$.


\subsection{Selection of the VIMF solutions}

\begin{figure}
\includegraphics[width=8.5cm]{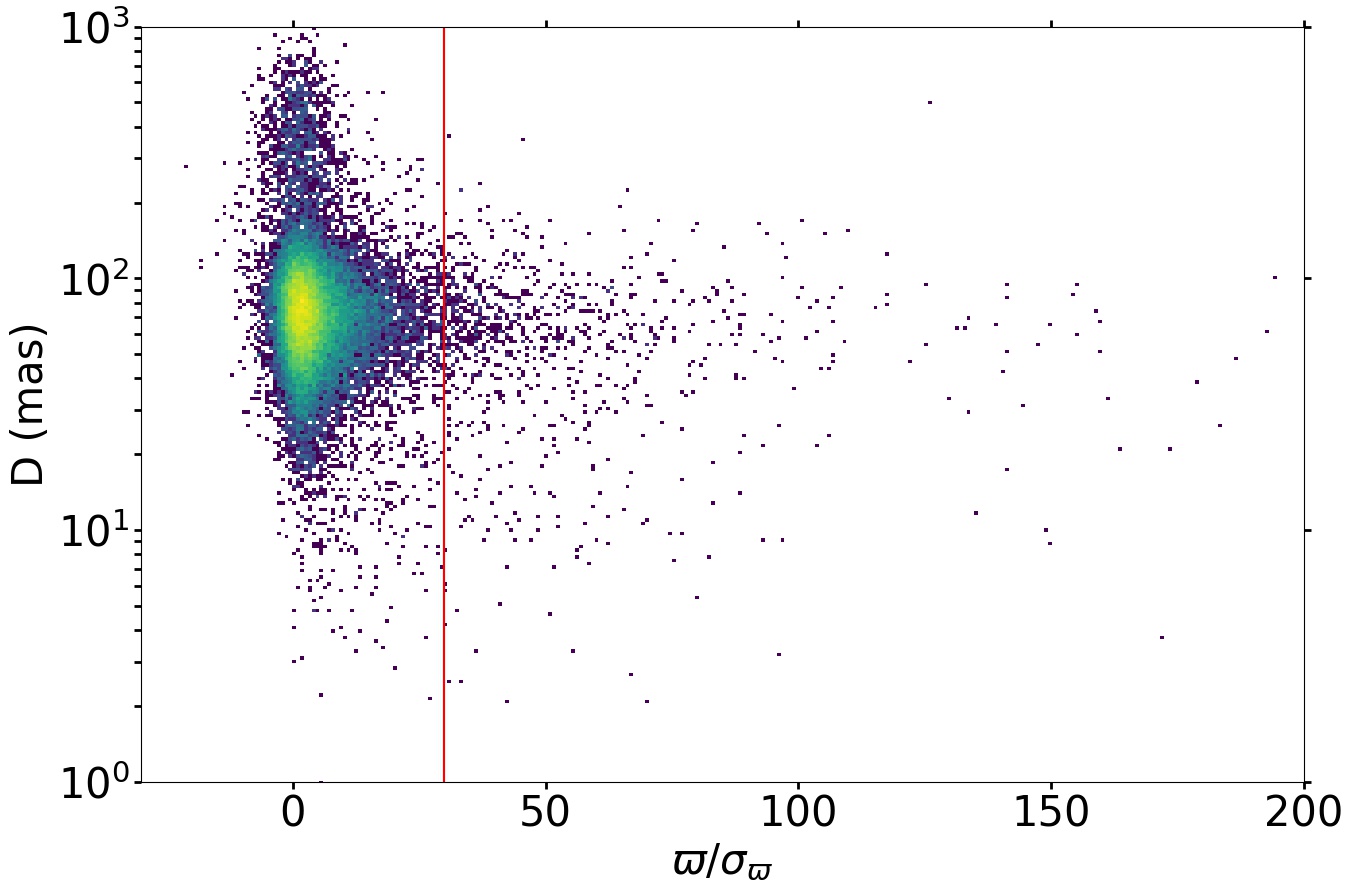}
\caption{The ``parallax significance'' vs. $D$ diagram of VIMF solutions obtained from a processing trial; most false solutions of
$D>100$~mas are eliminated if the selection criterion $\varpi/\sigma_\varpi>30$ (to the right of the vertical red line) is applied.
} 
\label{fig:VIMFtrial}
\end{figure}

The main criterion we have for estimating whether a VIMF solution is plausible is the apparent distance $D$ 
that separates the median photocentre from the variable component. As the resolution of {\it Gaia} is at least 180~mas 
\citep{Lindegren21} and the components of a VIM should never be separated, $D$ must remain under a limit of no 
more than a hundred mas; we are therefore looking here for quality criteria such that this condition is met.

The $F_2$ vs. $D$ and significance vs $D$ diagrams obtained after a first processing trial do not immediately show any limits in $F_2$
or significance that would allow almost only $D < 100$~mas. Fortunately, a selection becomes possible when we consider the ``parallax significance'', i.e. the quantity $\varpi/\sigma_\varpi$. It is clear from Fig.~\ref{fig:VIMFtrial} that the selection of solutions of parallax significance greater than 30
rejects most excessive values of $D$. Therefore, this criterion was applied in the main processing, in addition to the criteria $s>12$ and 
$F_2 < 25$. As before, the criteria for selecting alternative solutions were $s>5$ and $F_2$ less than that of the alternative solution that may have already been selected.

\begin{figure*}
\begin{subfigure}{.5\textwidth}
  \centering
  \includegraphics[width=0.9\linewidth]{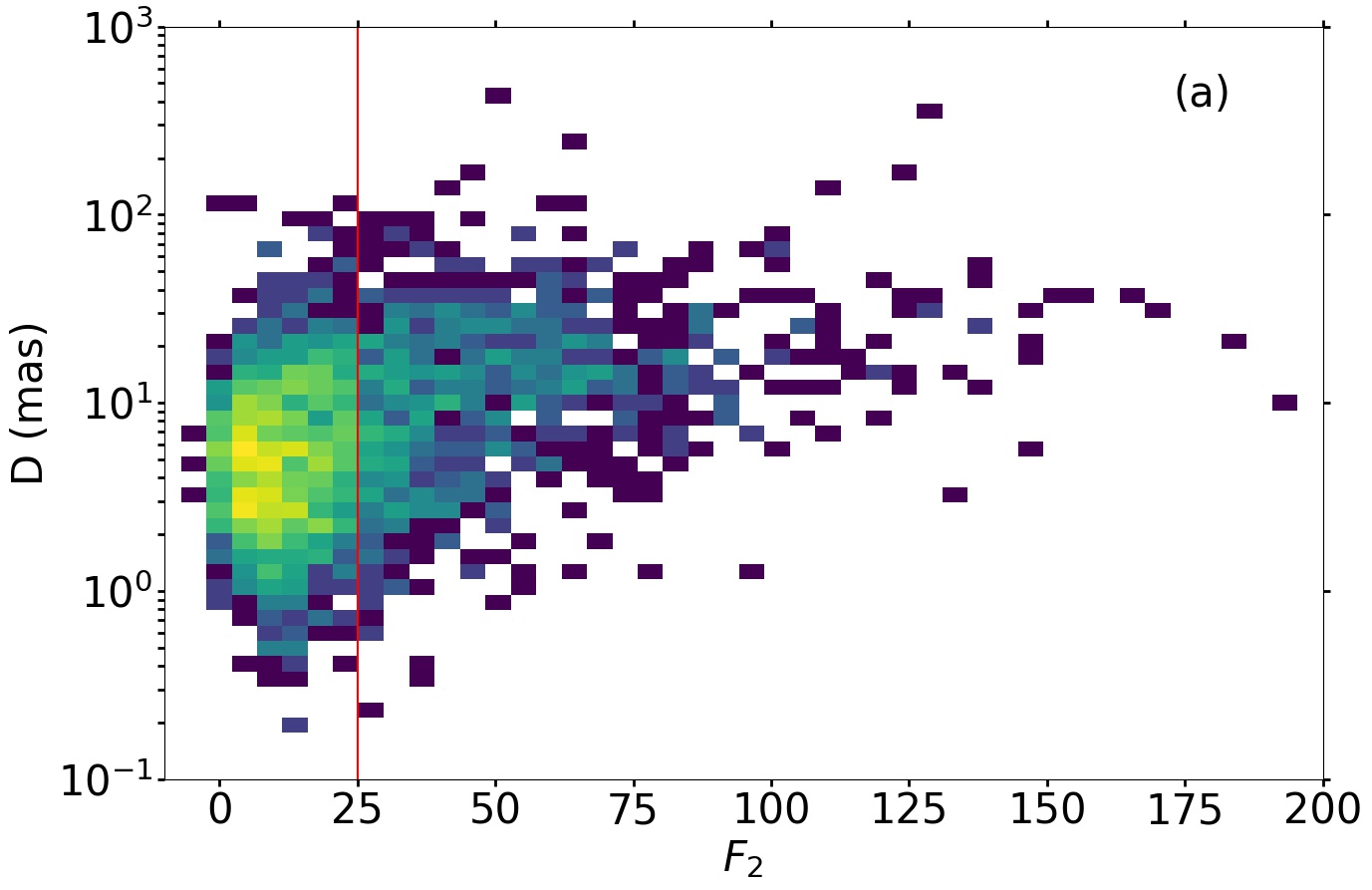}
\end{subfigure}
\begin{subfigure}{.5\textwidth}
  \centering
    \includegraphics[width=0.9\linewidth]{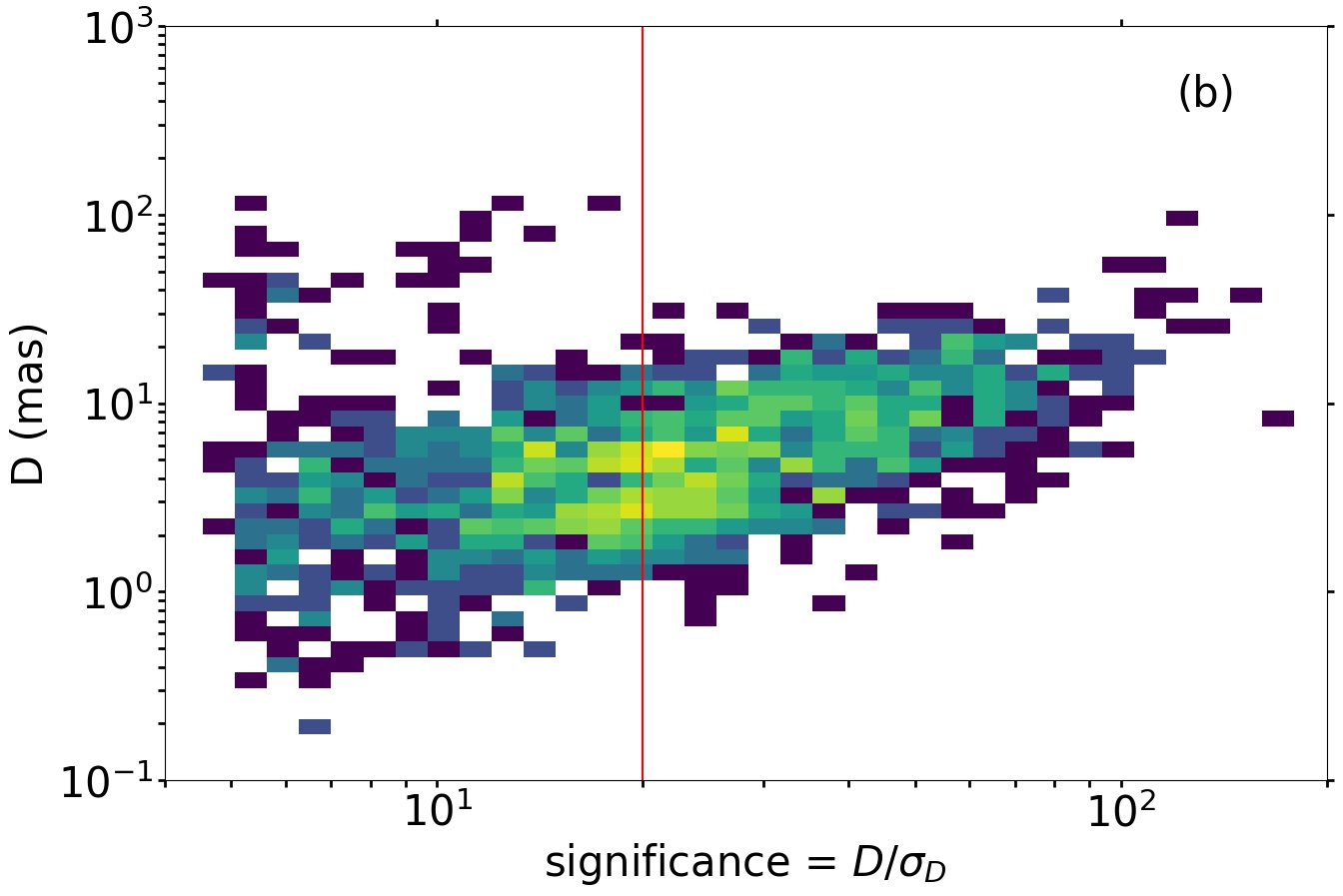}
\end{subfigure}
\caption{Post-processing filtering of the VIMF solutions. Panel (a): The $F_2$ vs. $D$ diagram of the 2508 solutions
obtained from the main processing. The proportion of large $D$, and therefore false 
solutions, increases gradually beyond $F_2=25$, to the right of the vertical red line. Panel (b): The significance vs. $D$ diagram of the 1660
solutions with $F_2<25$; acceptable solutions cluster on a sequence of $D$ increasing with significance;
abnormally large $D$ solutions are rejected by restricting to significances greater than 20 (to the right of the vertical red line),
leaving a final selection of 869 solutions.
}
\label{fig:VIMF}
\end{figure*}

Two thousand five hundred and eight VIMF solutions were obtained from the main processing.
The $F_2$ vs. $D$ diagram of these stars is shown in Fig.~\ref{fig:VIMF}a. As for the previous models, solutions with 
$F_2$ higher than 25 seem doubtful, because the proportion of large $D$ values in them increases, and, even more so, small $D$ values become increasingly rare as $F_2$ increases. They are therefore rejected, which leaves 
1660 solutions. The significance vs. $D$ diagram of these solutions shows a large scattering of $D$ values when significance is below 20, especially below 12.
However, the threshold was set at 20 as a precaution.
Beyond that, $D$ increases with $D/\sigma_D$ (Fig.~\ref{fig:VIMF}b), which corresponds to a value of $\sigma_D$ that is approximately 
constant and close to one tenth of mas, which seems acceptable. 

The 869 VIMF solutions thus retained in the end have distances $D$ 
below 40 mas, with three exceptions: three solutions of significance close to 100, when $D$ is 55, 56 and 101~mas.
These values are not unrealistic, and it seems possible that all VIMF solutions are valid.
Unfortunately, this may also be due to the low effectiveness of using $D$ to point out dubious solutions, as the application 
of other VIM models has shown below.

\subsection{The other VIM models}

As explained at the beginning of this section, four other VIM models were tried, from VIML to VIMO. Contrarely to the VIMF model,
the parameters of these models allow the estimation of a minimum value for the mass of the systems. The result was devastating:
for all these models, the minimum masses are concentrated between 10 and $10\;000$~${\cal M}_\sun$. We therefore decided to remove 
these models from the mass processing cascade.


\section{Discussion and conclusion}
\label{sect:discussion}

This first publication of astrometric binaries of different types observed by {\it Gaia} was made under difficult conditions, 
since the uncertainties of the measurements were often very underestimated. However, we were able to calculate hundreds of
thousands of reliable
solutions by keeping the weights of the observations and correcting the uncertainties. Nevertheless, this was done at the
cost of a drastic selection, based on very high quality criteria, and it would be difficult to use the selected solutions to deduce the statistical properties of binary stars, as the sample is incomplete and biased.

Because of the cascade structure that leads to the testing of different models, we may have retained solutions by direct 
acceptance that would have been superseded by another model, had it been tried. 
This choice has the advantage of preserving the quality of the orbital solutions, by assigning acceleration solutions to partially resolved double stars that would have been processed despite the filters set up to reject them. However, it was done at the expense of the completeness of the orbital solutions, and it is likely that some acceleration solutions actually correspond to binaries with periods short enough to have an orbital solution. 
Moreover, the quality criteria alone could not bring the acceleration solutions to a reasonable rate of high acceleration.
Ulrich Bastian studied this problem long after the post-processing was done, and he estimated that constant acceleration solutions are only highly reliable when the parallax is greater than 5~mas.
This criterion leaves only $15\,000$ solutions, but none with acceleration greater than 3 au.yr$^{-2}$. When only quality criteria are applied, ie. when criterion (1) of Sect.~\ref{subsect:selAcc} is not applied, the restriction to parallaxes greater than 5~mas reduces the $\Gamma>3$~au.yr$^{-2}$ rate from 3.1~\% to less than 1~\%; this confirms the effectiveness of this extremely drastic selection.
For future DRs, it is hoped that the treatment of partially resolved double stars, as well as testing all astrometric models and choosing the one most compatible with the data, will allow a wider selection of acceleration solutions.

The selection of orbital solutions was particularly severe to reject orbits that are concentrated in certain periods, which are likely to be produced by the satellite's scanning law. Despite these difficulties, about $165\,500$ astrometric orbits were finally selected in DR3.
The periods of the false orbits should be different in the following data releases, as the 
precession spin has been reversed during the last 6 months of observations that will be part of the DR4;
however, if these noise-induced orbits do not disappear at this time, they could be more difficult to detect, being distributed over a larger number of periods.
Furthermore, the lack of one-year periods, which comes from a degeneracy between parallactic and orbital motion, is expected to remain.

The acceleration and orbital models apply to unresolved binaries with either components that do not vary photometrically, or one component always much brighter than the other. This limits their interest in the study of young stars, or more generally, pulsating stars. 
VIM models extend the scientific scope of astrometric binaries, provided that only one of the components is photometrically variable.
Unfortunately,
the number of VIM solutions appears to be very modest, numbering only a few hundred, limited to the VIMs with fixed components.
This rarity is partly due to the fact that the model has only been tried at 
the end of the cascade, and we do not know how many acceptable VIM solutions were lost due to the acceptance of a solution of another type. 

When both components are variable, the VIM solution can be very degraded: When the brightnesses of the two stars vary in the same direction, the displacement 
of the photocentre can be very small while the brightness of the system varies greatly; for the VIMF model, this is interpreted as a short D distance.
Conversely, if the brightness of one component increases while that of the other decreases, the fluctuation in total brightness that may be small while the 
photocentre displacement is large. As the luminosities of the components do not generally vary in phase, one can expect VIMF solutions of large $\chi^2$, and therefore large $F_2$, with $D$ values that can be minimal or, on the contrary, very large. A large $\chi^2$ leads to a decrease in significance, thanks to the coefficient $c$ of Eq.~\ref{eq:corsig}, and, as a result, the pollution of VIMF solutions by systems whose two components are photometrically variable will be limited thanks to the high acceptance thresholds that have been adopted. 

Another difference between fixed-component VIMs and other solutions is that
the selection could only be adjusted on the basis of the median distance between the photocentre and
the photometric variable component, excluding any dynamic criteria. Therefore, these solutions should be considered with 
caution. 
In turn, it is possible that some acceleration or orbital solutions have been assigned to objects that should have received a VIM solution.
However, these cases are too rare to constitute detectable pollution: \cite{DPACP-100} have shown that, statistically, acceleration solutions have proper 
motions affected by orbital motion, which would not be the case for VIMFs. Similarly, if VIMFs receive an orbital solution, the inclination of the 
solution must be 90 degrees, and the distribution of inclinations does not show an excess for this value.

Despite these limitations raised above, the DR3 is a major advance in the field of binary stars. The hundreds of thousands of new 
astrometric orbits will have considerable importance in the future, compared to the less than 3000 visual binary star orbits 
in the Sixth Catalog of Orbits of Visual Binary Stars\footnote{https://www.usno.navy.mil/USNO/astrometry/optical-IR-prod/wds/orb6}.
Such a change of scale could open a new era in the statistical study of binary stars, especially if the next DRs could be 
less affected by selection effects. Instead of looking at the properties
of pairs with a primary component of a given type, it may be possible to consider pairs with a given total mass. Such an 
approach would shed new light on the star formation process, provided that the many biases affecting binary selection are taken into account. 


\begin{acknowledgements} 

This work is part of the reduction of the {\it Gaia} satellite observations (https://www.cosmos.esa.int/gaia).
The {\it Gaia} space mission is operated by the European Space Agency, and the
data are being processed by the {\it Gaia} Data Processing and Analysis Consortium (DPAC, https://www.cosmos.esa.int/web/gaia/dpac/consortium)). The {\it Gaia} archive website is https://archives.esac.esa.int/gaia. Funding for the DPAC is provided by national institutions, in particular the institutions participating in the {\it Gaia} MultiLateral Agreement (MLA). We acknowledge for the financial support of the french ``Centre National d'études spatiales'' (CNES)
and of the BELgian federal Science Policy Office (BELSPO) through a PROgramme de Développement d'Expériences scientifiques (PRODEX).
Over the two decades that it took to develop the calculation methods and write the software, we benefited from the contributions of Sylvie Jancart, Rémy Onsay, Didier Pelat and Patrick Gempin. We also received constant support from François Mignard and Alain Jorissen. The acceptability of acceleration solutions was discussed with Ulrich Bastian and Lennart Lindegren, using data produced by Sergei Klioner and Hagen Steidelm\"uller.
Finally, we are grateful to Michael Biermann and to an anonymous referee for their careful reading of the manuscript and the relevance of their comments.

\end{acknowledgements}




\begin{appendix}

\section{Conversion of Thiele-Innes elements ($A,B,F,G$) into Campbell elements ($a,i,\Omega,\omega$) and calculation of the uncertainties}
\label{Sect.ABFG2Campbell}

The Thiele-Innes (TI) elements, $A$, $B$, $F$ and $G$ are given by the following equations \citep{Binnendijk, Heintz}:

\begin{equation}
\begin{array}{ll}
A = & a \; (\cos \omega \cos \Omega - \sin \omega \sin \Omega \cos i) \\
B = & a \; (\cos \omega \sin \Omega + \sin \omega \cos \Omega \cos i) \\
F = &-a \; (\sin \omega \cos \Omega + \cos \omega \sin \Omega \cos i) \\
G = & -a \; (\sin \omega \sin \Omega - \cos \omega \cos \Omega \cos i) 
\end{array}
\label{eq:ABFG}
\end{equation} 

\noindent
where $a$, $i$, $\Omega$ and $\omega$ are the Campbell elements. The definitions of these elements and
their calculations from $A$, $B$, $F$, $G$ are given hereafter:

\begin{itemize}

\item $a$ (or $a_0$ for a photocentric orbit) is the semi-major axis of the orbit, in the same unit as the TI elements.

\item $i$ is the inclination of the orbit, i.e. the angle between the orbital plane and the plane of the sky.
$i \in [0, \pi/2]$ when the orbital motion is in the direct sense, and $i \in [\pi/2, \pi]$ when the photocentre is revolving
around the barycentre in a clockwise direction.

\item $\Omega$ is the position angle of the ascending node. For us, the ascending node is the intersection between
the orbital plane and the plane of the sky in the direction where the right ascension is increasing. Therefore,
$\Omega \in [0, \pi]$.

\item $\omega$ is the periastron longitude, measured in the orbital plane from the ascending node defined above.
$\omega$ is measured in the sense ofthe orbital motion.

\end{itemize}


\citet{Binnendijk} gives the following method for converting TI elements into Campbell elements:

The semi-major axis $a$, is derived from:

\begin{equation}
\begin{array}{ll}
u = & (A^2 + B^2 + F^2 + G^2) / 2 \\
v = & AG - BF \\
a = & \sqrt{u + \sqrt{(u+v)(u-v)}}
\end{array}
\label{eq:a}
\end{equation}


The angles $\Omega$ and $\omega$ are derived simultaneously. Preliminary estimates, between
$-\pi$ and $\pi$, are derived as follows:

\begin{equation}
\begin{array}{ll}
\omega+\Omega = & \arctan \frac {B-F}{A+G} \pmod \pi \\
 & \\
\omega-\Omega = & \arctan \frac {B+F}{G-A} \pmod \pi
\end{array}
\label{eq:opmO}
\end{equation}

The exact values of $\omega+\Omega$ are obtained, modulo $2\pi$,
taking into account
that $\sin (\omega+\Omega)$ and $(B-F)$ have the same sign.
Similarly, $\sin (\omega-\Omega)$ has the same sign as $(-B-F)$.
$\omega$ and $\Omega$ are then derived from the equations:

\begin{equation}
\begin{array}{ll}
\omega = & \frac{(\omega+\Omega) + (\omega-\Omega)}{2} \pmod \pi \\
 & \\
\Omega = & \frac{(\omega+\Omega) - (\omega-\Omega)}{2} \pmod \pi
\end{array}
\label{eq:oOmega}
\end{equation}

When $\Omega$ is negative, $\pi$ must still be added to both $\Omega$ and
$\omega$. After that, a correction of $2\pi$ may still be applied in order
to meet the conditions $\Omega \in [0, \pi]$ and $\omega \in [0, 2\pi]$.


To derive the inclination $i$, we first need to calculate two intermediate terms,
called $d_1$ and $d_2$ hereafter:

\begin{equation}
\begin{array}{ll}
d_1 = & \| (A + G) \cos(\omega-\Omega) \| \\
d_2 = & \| (F - B) \sin(\omega-\Omega) \|
\end{array}
\label{eq:d1d2}
\end{equation}

In order to obtain the value of $i$, the calculation is different
depending on whether $d_1$ is greater than $d_2$, or the reverse:

\begin{equation}
\begin{array}{ll}
\mathrm{if} \;(d_1 \geq d_2), & i = 2 \arctan \sqrt{ \| (A-G) \cos(\omega+\Omega) \| \; / \; d_1} \\
\mathrm{else,}             & i = 2 \arctan \sqrt{ \| (B+F) \sin(\omega+\Omega) \| \; / \; d_2} 
\end{array}
\label{eq:inc} 
\end{equation}


The uncertainties of $a$, $i$, $\Omega$ and $\omega$ are derived by differentiating the
equations above, and by taking into account error propagation. 

We call hereafter $\sigma_p$ the error on element $p$ and $\mathrm{cov}$(X,Y) the term of the
variance covariance matrix of the solution linking the elements $X$ and $Y$, ie 
$\mathrm{cov}(X,Y) = \rho_{XY} \sigma_X \sigma_Y$, 
where $\rho_{XY}$ is the correlation coefficient between $X$ and $Y$. The uncertainties are
then given by the equations below.

To derive the uncertainty of $a$, the following intermediate terms must be calculated
from $u$ and $v$ derived in Eq.~\ref{eq:a}:

\begin{equation}
\begin{array}{ll}
t_A = & A  \; + \; (A \; u - G \; v) \; /\sqrt{u^2 - v^2} \\
t_B = & B  \; + \; (B \; u + F \; v) \; /\sqrt{u^2 - v^2} \\
t_F = & F  \; + \; (F \; u + B \; v) \; /\sqrt{u^2 - v^2} \\
t_G = & G  \; + \; (G \; u - A \; v) \; /\sqrt{u^2 - v^2} 
\end{array}
\label{eq:tABFG} 
\end{equation}

The uncertainty of the semi-major axis is then:

\begin{equation}
\begin{array}{ll}
\sigma_a = \frac{1}{2a} \; \times \; [ & t_A^2 \sigma_A^2 + t_B^2 \sigma_B^2 + t_F^2 \sigma_F^2 + t_G^2 \sigma_G^2 \\
 & + \; 2 t_A t_B \mathrm{cov}(A,B) + 2 t_A t_F \mathrm{cov}(A,F)   \\
 & + \; 2 t_A t_G \mathrm{cov}(A,G) + 2 t_B t_F \mathrm{cov}(B,F) \\
 & + \; 2 t_B t_G \mathrm{cov}(B,G) + 2 t_F t_G \mathrm{cov}(F,G) \; ]^{1/2}
\end{array}
\label{eq:siga}
\end{equation}


The uncertainties of $\omega$ and $\Omega$ are also derived using dedicated intermediate terms.
We first calculate the followings:

\begin{equation}
\begin{array}{ll}
k = & (A+G)^2 + (B-F)^2 \\
l = & (G-A)^2 + (B+F)^2 
\end{array}
\label{eq:kl}
\end{equation}

These terms are used to derive other intermediate terms dedicated to the calculation of the
uncertainty of $\omega$:

\begin{equation}
\begin{array}{ll}
w_A = & (F-B)/k + (B+F)/l \\
w_B = & (A+G)/k + (G-A)/l \\
w_F = & -(A+G)/k + (G-A)/l \\
w_G = & (F-B)/k - (B+F)/l
\end{array}
\label{eq:wABFG}
\end{equation}
 
The uncertainty of $\omega$, $\sigma_\omega$ is then derived with the equation hereafter:

\begin{equation}
\begin{array}{ll}
 \sigma_\omega = & [ w_A^2 \sigma_A^2 + w_B^2 \sigma_B^2 + w_F^2 \sigma_F^2 + w_G^2 \sigma_G^2 \\
       &        + \; 2 \;   ( \; w_A w_B \;  \mathrm{cov}(A,B) +   w_A w_F \;  \mathrm{cov}(A,F)  \\
       &        + w_A w_G \;  \mathrm{cov}(A,G) + w_B w_F \;  \mathrm{cov}(B,F) \\
       &        + w_B w_G \;  \mathrm{cov}(B,G) + w_F w_G \;  \mathrm{cov}(F,G) \; ) \; ]^{1/2} \; / \; 2
\end{array}
\label{eq:sigw}
\end{equation}

The terms from Eq.~\ref{eq:kl} are also used to compute the following intermediate terms, which are
required to derive the uncertainty of $\Omega$:

\begin{equation}
\begin{array}{ll}
O_A = & (F-B)/k - (B+F)/l \\
O_B = & (A+G)/k - (G-A)/l \\
O_F = & -(A+G)/k - (G-A)/l \\
O_G = & (F-B)/k + (B+F)/l
\end{array}
\label{eq:OABFG}
\end{equation}

The uncertainty of $\Omega$ is then:

\begin{equation}
\begin{array}{ll}
 \sigma_\Omega = & [ O_A^2 \sigma_A^2 + O_B^2 \sigma_B^2 + O_F^2 \sigma_F^2 + O_G^2 \sigma_G^2 \\
     &          + \; 2 \; ( \; O_A  O_B \;  \mathrm{cov}(A,B) + O_A  O_F \;  \mathrm{cov}(A,F)   \\
     &          + O_A  O_G \;  \mathrm{cov}(A,G) + O_B  O_F \;  \mathrm{cov}(B,F)               \\
     &          + O_B  O_G \;  \mathrm{cov}(B,G) + O_F  O_G \;  \mathrm{cov}(F,G) \; ) \; ]^{1/2} \; / \; 2
\end{array}
\label{eq:sigOmega}
\end{equation}

Like the derivation of $i$, the derivation of the uncertainty of the inclination 
depends on a comparison between $d_1$ and $d_2$, derived in Eq.~\ref{eq:d1d2} above.

$\sigma_i$ is derived as follows. We first calculate the
terms:

\begin{equation}
\begin{array}{ll}
q = & \sin \Omega \cos \Omega \\
r = & \sin \omega \cos \omega \\
\end{array}
\label{eq:qr}
\end{equation}

if $d_1 \geq d_2$, the continuation of the calculation is as explained hereafter.
An extra term, $p_1$, is added to $q$ and $r$:

\begin{equation}
p_1 =  \cos(\omega-\Omega) \; \cos(\omega+\Omega) 
\label{eq:p1}
\end{equation}

The intermediate terms dedicated to the calculation of $\sigma_i$ are then:

\begin{equation}
\begin{array}{ll}
h_A = & 2 G p_1 + (G^2 - A^2) \; (q w_A  + r O_A ) \\
h_B = & (G^2 - A^2) \; (q w_B + r O_B) \\
h_F = & (G^2 - A^2) \; (q w_F + r O_F) \\
h_G = & -2 A p_1 + (G^2 - A^2) \; (q w_G + r O_G)
\end{array}
\label{eq:hABFG}
\end{equation}

and, still for $d_1 \geq d_2$, $\sigma_i$ is derived from the equation:

\begin{equation}
\begin{array}{ll}
\sigma_i = & [ h_A^2 \sigma_A^2 + h_B^2 \sigma_B^2 + h_F^2 \sigma_F^2 + h_G^2 \sigma_G^2 \\
  &          + \; 2 \; ( \; h_A h_B  \; \mathrm{cov}(A,B) +  h_A h_F  \; \mathrm{cov}(A,F) \\
  &          + h_A h_G  \; \mathrm{cov}(A,G) + h_B h_F  \; \mathrm{cov}(B,F)     \\
  &          + h_B h_G  \; \mathrm{cov}(B,G) + h_F h_G  \; \mathrm{cov}(F,G) \; ) \; ]^{1/2} \\
  &          / \; [ \; ( \tan (i/2) + \tan^3 (i/2)) \;  d_1^2 \; ]
\end{array} 
\label{eq:sigi1}
\end{equation}

When $d_2 < d_1$, the additional term $p_2$ is derived:

\begin{equation}
p_2 =  \sin(\omega-\Omega) \; \sin(\omega+\Omega) 
\label{eq:p2}
\end{equation}

The intermediate terms dedicated the calculation of $\sigma_i$ are now:

\begin{equation}
\begin{array}{ll}
g_A = &  (B^2 - F^2) \; ( q w_A - r O_A) \\
g_B = & 2 F p_2 + (B^2 - F^2) \; (q w_B - r O_B \\
g_F = & -2 B p_2 + (B^2 - F^2) \; (q w_F - r O_F)  \\
g_G = &   (B^2 - F^2) \; (q w_G - r O_G)
\end{array}
\label{eq:gABFG}
\end{equation}

and, still for $d_1 < d_2$, $\sigma_i$ is derived from the equation:

\begin{equation}
\begin{array}{ll}
\sigma_i = & [ g_A^2 \sigma_A^2 + g_B^2 \sigma_B^2 + g_F^2 \sigma_F^2 + g_G^2 \sigma_G^2 \\
  &          + \; 2 \; ( \; g_A g_B  \; \mathrm{cov}(A,B) +  g_A g_F  \; \mathrm{cov}(A,F) \\
  &          + g_A g_G  \; \mathrm{cov}(A,G) + g_B g_F  \; \mathrm{cov}(B,F)     \\
  &          + g_B g_G  \; \mathrm{cov}(B,G) + g_F g_G  \; \mathrm{cov}(F,G) \; ) \; ]^{1/2} \\
  &          / \; [ \; ( \tan (i/2) + \tan^3 (i/2)) \;  d_2^2 \; ]
\end{array} 
\label{eq:sigi2}
\end{equation}

The four Campbell elements and their uncertainties are thus deduced from the solutions calculated in TI elements.
However, it is not possible to estimate the correlation coefficients between the terms.


\section{Conversion of Thiele-Innes elements ($C_1,H_1$) into elements (${\rm a}_1,\omega_1$) and calculation of the uncertainties}
\label{Sect.CH2Campbell}

The $C_1$ and $H_1$ elements are functions of the ${\rm a}_1$ and $\omega_1$ elements that come from the spectroscopic
orbit:

\begin{equation}
\begin{array}{ll}
C_1 = & {\rm a}_1 \sin i \sin \omega_1 \\
H_1 = & {\rm a}_1 \sin i \cos \omega_1
\end{array} 
\label{eq:C1H1}
\end{equation}

from which it appears that:

\begin{equation}
\begin{array}{ll}
{\rm a}_1 = &  \sqrt{C_1^2 + H_1^2} / \sin i  \\
\omega_1 =  & \arctan C_1/H_1
\end{array} 
\label{eq:a1w1}
\end{equation}

\noindent
knowing that sign$(\sin \omega_1)={\rm sign}(C_1) \textrm{ and sign}(\cos \omega_1)={\rm sign}(H_1)$.

$\omega_1$ must be approximately equal to either $\omega$, or $\omega + \pi$. This ambiguity is due to differences in definition: $\omega$ is the longitude of the periastron of the photocentre orbit, which, in the absence of radial velocity, is measured from the node whose position angle is between 0 and $\pi$. For its part, $\omega_1$ is the longitude of the periastron of the orbit of the spectroscopically most visible component, measured from the node where this component moves away from the Sun. Therefore, the true orientation of the orbital plane will depend on the position of the photocentre with respect to to the barycentre and the spectroscopic component: if the photocentre is assumed to be between the barycentre and the spectroscopic component, then the true ascending node and periastron of the astrometric orbit are marked by the same angles as those of the spectroscopic orbit. If $\omega$ and $\omega_1$ are significantly different, it is necessary to add $\pm \pi$ to $\omega$ to make them roughly coincide, and also to add $\pi$ to $\Omega$ to take into account the change of reference node. 

Alternatively, the photocentre may be opposite to the spectroscopic component (this may occur, for example, when the spectral type of the brightest component is very different from the templates used for spectroscopic reduction, so that the spectroscopic orbit corresponds to the least bright binary component). When this occurs, $\omega$ must be approximately equal to $\omega_1 \pm \pi$. If it is not, $\pm \pi$ must be added to $\omega$ and $\pi$ must be added to $\Omega$ for the orientation of the astrometric orbit to be completely defined.

The calculation of the uncertainty of ${\rm a}_1$ depends on the values of $d_1$ and $d_2$, given by Eq.~\ref{eq:d1d2}. When $d_1 \ge d_2$,
we compute the following terms:

\begin{equation}
\left(
\begin{array}{l}
q_A \\
q_B \\
q_F \\
q_G 
\end{array}
\right) =
\left(
\begin{array}{l}
h_A \\
h_B \\
h_F \\
h_G 
\end{array}
\right) \times
\frac{{\rm a}_1}{\tan i \tan \frac{i}{2} \left( 1+\tan^2 \frac{i}{2} \right) d_1^2}
\label{eq:qABFG}
\end{equation}

\noindent
where $h_A$, $h_B$, $h_F$, and $h_G$ are taken from Eq.~\ref{eq:hABFG}, and

\begin{equation}
\left(
\begin{array}{l}
q_C \\
q_H 
\end{array}
\right) =
\left(
\begin{array}{l}
C_1 \\
H_1 
\end{array}
\right) \times
\frac{{\rm a}_1}{C_1^2 + H_1^2}
\label{eq:qCH}
\end{equation}

The uncertainty $\sigma_{{\rm a}_1}$ is then calculated with equation:

\begin{equation}
\begin{array}{ll}
\sigma_{{\rm a}_1} = & [q_A^2 \sigma_A^2 + q_B^2 \sigma_B^2 + q_F^2 \sigma_F^2 + q_G^2 \sigma_G^2  + q_C^2 \sigma_{C_1}^2 + q_H^2 \sigma_{H_1}^2 \\
  &          + \; 2 \; ( \; q_A q_B  \; \mathrm{cov}(A,B) +  q_A q_F  \; \mathrm{cov}(A,F) \\
  &          + q_A q_G  \; \mathrm{cov}(A,G)   + q_A q_C  \; \mathrm{cov}(A,C_1)  \\
  &          + q_A q_H  \; \mathrm{cov}(A,H_1) + q_B q_F  \; \mathrm{cov}(B,F)     \\
  &          + q_B q_G  \; \mathrm{cov}(B,G)   + q_B q_C  \; \mathrm{cov}(B,C_1)  \\
  &          + q_B q_H  \; \mathrm{cov}(B,H_1) + q_F q_G  \; \mathrm{cov}(F,G)     \\
  &          + q_F q_C  \; \mathrm{cov}(F,C_1) + q_F q_H  \; \mathrm{cov}(F,H_1) \\
  &          + q_G q_C  \; \mathrm{cov}(G,C_1) + q_G q_H  \; \mathrm{cov}(G,H_1) \\
  &          + q_C q_H  \; \mathrm{cov}(C_1,H_1) \; ) \; ]^{1/2} 
\end{array}
\label{eq:siga11}
\end{equation}

When $d_2 > d_1$, these equations become:

\begin{equation}
\left(
\begin{array}{l}
r_A \\
r_B \\
r_F \\
r_G 
\end{array}
\right) =
\left(
\begin{array}{l}
g_A \\
g_B \\
g_F \\
g_G 
\end{array}
\right) \times
\frac{{\rm a}_1}{\tan i \tan \frac{i}{2} \left( 1+\tan^2 \frac{i}{2} \right) d_2^2}
\label{eq:rABFG}
\end{equation}

\noindent
where $g_A$, $g_B$, $g_F$, and $g_G$ are taken from Eq.~\ref{eq:gABFG}, and

\begin{equation}
\left(
\begin{array}{l}
r_C \\
r_H 
\end{array}
\right) =
\left(
\begin{array}{l}
C_1 \\
H_1 
\end{array}
\right) \times
\frac{{\rm a}_1}{C_1^2 + H_1^2}
\label{eq:rCH}
\end{equation}

The uncertainty $\sigma_{{\rm a}_1}$ is then calculated with the equation:

\begin{equation}
\begin{array}{ll}
\sigma_{{\rm a}_1} = & [r_A^2 \sigma_A^2 + r_B^2 \sigma_B^2 + r_F^2 \sigma_F^2 + r_G^2 \sigma_G^2  + r_C^2 \sigma_{C_1}^2 + r_H^2 \sigma_{H_1}^2 \\
  &          + \; 2 \; ( \; r_A r_B  \; \mathrm{cov}(A,B) +  r_A r_F  \; \mathrm{cov}(A,F) \\
  &          + r_A r_G  \; \mathrm{cov}(A,G)   + r_A r_C  \; \mathrm{cov}(A,C_1)  \\
  &          + r_A r_H  \; \mathrm{cov}(A,H_1) + r_B r_F  \; \mathrm{cov}(B,F)     \\
  &          + r_B r_G  \; \mathrm{cov}(B,G)   + r_B r_C  \; \mathrm{cov}(B,C_1)  \\
  &          + r_B r_H  \; \mathrm{cov}(B,H_1) + r_F r_G  \; \mathrm{cov}(F,G)     \\
  &          + r_F r_C  \; \mathrm{cov}(F,C_1) + r_F r_H  \; \mathrm{cov}(F,H_1) \\
  &          + r_G r_C  \; \mathrm{cov}(G,C_1) + r_G r_H  \; \mathrm{cov}(G,H_1) \\
  &          + r_C r_H  \; \mathrm{cov}(C_1,H_1) \; ) \; ]^{1/2} 
\end{array}
\label{eq:siga12}
\end{equation}

The uncertainty of $\omega_1$, $\sigma_{\omega_1}$, is derived from:

\begin{equation}
\sigma_{\omega_1} = \frac{\cos^2 \omega_1}{H_1^2} [H_1^2 \sigma_{C_1}^2 + C_1^2 \sigma_{H_1}^2 - 2 C_1 H_1 \; \mathrm{cov}(C_1,H_1)]^{1/2}
\label{eq:sigomega1}
\end{equation}

\end{appendix}


\end{document}